\documentclass[journal]{IEEEtran}

\ifCLASSINFOpdf
\else
\fi
\usepackage{cite}

\usepackage[cmex10]{amsmath}

\usepackage[caption=false,font=footnotesize]{subfig}

\usepackage{stfloats}

\usepackage{graphicx}
\usepackage{color}
\usepackage{placeins}
\usepackage{float}
\usepackage[colorlinks=true,linkcolor=black,citecolor=black,urlcolor=black]{hyperref}
\usepackage{tabularx,colortbl,multirow}
\usepackage{enumerate}
\usepackage{enumitem}
\usepackage{bm}
\newcommand\figref[1]{Fig.~\ref{#1}}

\renewcommand\eqref[1]{(\ref{#1})}

\usepackage{tikz}
\usetikzlibrary{shapes,decorations,decorations.pathreplacing}
\usetikzlibrary{arrows,shapes,positioning,calc,fadings,patterns}
\usetikzlibrary{decorations,decorations.pathreplacing,decorations.pathmorphing}
\usepackage{tikz-3dplot}

\usepackage{booktabs}     % used for the table
\usepackage{threeparttable}
\usepackage{amssymb}

\usepackage{algorithm}
\usepackage{algcompatible}
\usepackage{amsmath}

\newcommand{\re}[1]{\textcolor{black}{#1}}

\begin{document}
	\title{
		Physics-Based Trajectory Design for Cellular-Connected~UAV~in~Rainy~Environments Based on Deep Reinforcement Learning
		%\vspace*{-0.1cm} % if need to reduce space in between
	}
	
	\author{Hao~Qin$^1$,~Zhaozhou~Wu$^1$,~\IEEEmembership{Student Member,~IEEE}, and~Xingqi~Zhang$^{1, 2}$,~\IEEEmembership{Senior Member,~IEEE}
		\thanks{Manuscript received XX XX, XXXX; revised XX XX, XXXX; accepted XX XX, XXXX.}
		\thanks{$^1$The authors are with the School of Electrical and Electronic Engineering, University College Dublin, Ireland (e-mail: hao.qin@ucdconnect.ie, zhaozhou.wu@ucdconnect.ie, xingqi.zhang@ucd.ie).}
		\thanks{$^2$The author is with the Department of Electrical and Computer Engineering, University of Alberta, Canada (e-mail: xingqi.zhang@ualberta.ca).}% <-this % stops a space
		\thanks{Color versions of one or more of the figures in this paper are available online at http://ieeexplore.ieee.org.}
		\thanks{Digital Object Identifier: XXXX}
		%\vspace*{-0.1cm}
	}
	
	% journal name
	%\markboth{IEEE ANTENNAS AND WIRELESS PROPAGATION LETTERS}%
	\markboth{IEEE TRANSACTIONS ON WIRELESS COMMUNICATIONS}%
	{ZHANG \textit{et al} \MakeLowercase{\textit{}}: }
	

	\maketitle
	
	\vspace{-0.5cm}
	
	\begin{abstract}
		Cellular-connected unmanned aerial vehicles~(UAVs) have gained increasing attention due to their potential to enhance conventional UAV capabilities by leveraging existing cellular infrastructure for reliable communications between UAVs and base stations. They have been used for various applications, including weather forecasting and search and rescue operations. However, under extreme weather conditions such as rainfall, it is challenging for the trajectory design of cellular UAVs, due to weak coverage regions in the sky, limitations of UAV flying time, and signal attenuation caused by raindrops. To this end, this paper proposes a physics-based trajectory design approach for cellular-connected UAVs in rainy environments. A physics-based electromagnetic simulator is utilized to take into account detailed environment information and the impact of rain on radio wave propagation. The trajectory optimization problem is formulated to jointly consider UAV flying time and signal-to-interference ratio, and is solved through a Markov decision process using deep reinforcement learning algorithms based on multi-step learning and double Q-learning. Optimal UAV trajectories are compared in examples with homogeneous atmosphere medium and rain medium. Additionally, a thorough study of varying weather conditions on trajectory design is provided, and the impact of weight coefficients in the problem formulation is discussed. The proposed approach has demonstrated great potential for UAV trajectory design under rainy weather conditions.
	\end{abstract}
	
	
	\IEEEpeerreviewmaketitle

	%%%%%%%%%%%%%%%%%%%%%%%%%%%%%%%%%%%%%%%%%%%%%%%%%%%%%%%%%%%%%%%%%%%%%%%%%%%%%%%%
	%%%%%%%%%%%%%%%%%%%%%%%%%%%%%%%%%%%%%%%%%%%%%%%%%%%%%%%%%%%%%%%%%%%%%%%%%%%%%%%%
	\section{Introduction}
	
	\IEEEPARstart{U}{nmanned} aerial vehicles (UAVs) are gaining increasing popularity in a variety of applications, such as surveillance, monitoring, and inspection. 
	However, the limited range of traditional UAVs can be a significant obstacle to their widespread adoption~\cite{Zeng19Accessing, Gupta15Survey, Yang2019energy, Khuwaja18, Azari18, Ding20UAV}.
	To overcome this challenge, cellular-connected UAVs have emerged as an innovative wireless technology that integrates UAVs into cellular networks~\cite{Amorim17, Fotouhi2019survey, Zhang2019cellular, Azari20, Zhan2020energy}.
	On one hand, dedicated UAVs can serve as communication relays or even aerial base stations (BSs) to provide wireless communications among BSs and users that are located at long distances or cannot be directly connected~\cite{Zeng16}.
	On the other hand, cellular networks can provide a reliable and secure communication link between UAVs and ground control stations. This allows operators to fly the UAVs over longer distances and in more challenging environments, without the need for visual contact. In particular, the UAVs are supported by the existing cellular BSs on the ground.
	
	In complex meteorological conditions such as rain, cellular-connected UAVs can be utilized for various applications, such as weather monitoring, disaster response, and agricultural inspections. 
	However, several critical challenges need to be addressed when integrating UAVs into existing cellular networks in rainy environments. Firstly, weak coverage regions exist in the sky because such BSs and conventional cellular networks are primarily designed to serve terrestrial user equipment (UE), and BS antennas are typically downtilted towards the ground. 
	Therefore, the trajectories of the cellular-connected UAVs need to be well-designed to meet the communication requirements and mission specifications. 
	Secondly, as aerial UEs, UAVs require efficient and site-specific trajectories, which necessitates the development of rapid and physics-based models that can extract detailed information about the surroundings and design an appropriate trajectory. 
	However, statistical models require time-consuming and labor-intensive measurement campaigns in rainy environments, while deterministic models can be employed as an alternative.
	%	Thirdly, UAVs have numerous applications in rainy environments, including weather forecasting, flood monitoring, agriculture, infrastructure inspection, and search and rescue operations.
	%	with the emergence of new wireless communication systems such as 5G/6G systems, more and more high-frequency wireless systems are being applied.
	Thirdly, under rainy weather conditions, wireless communication systems can experience significant degradation in performance, which can have a direct impact on the optimal trajectory of UAVs. This highlights the necessity for accurate wave propagation models in rainy environments.
	%	The impact of the wave propagation medium, such as rain, on wireless communication systems, should not be ignored. 
	\re{The current techniques on cellular-connected UAV path planning do not account for rainy weather conditions. Addressing this limitation is crucial to ensure the reliability and effectiveness of UAV communication in various weather conditions and to enhance the overall performance of cellular-connected UAV systems.}
	All of these factors motivate the development of physics-based trajectory design approach for cellular-connected UAVs in rainy environments, which is missing in the current literature. 
	%However, there is missing in the literature that an accurate trajectory design method for cellular UAVs in rainy environments.
	%	The existing literature lacks research about an accurate trajectory design method for cellular UAVs in rainy environments.
	
	\subsection{Related Prior Work}
	Trajectory design for cellular-connected UAVs has been extensively investigated in recent years~\cite{Wu18joint, Zhan2022energy, Chen2021joint, Hu2022trajectory, Zhang2019trajectory}.
	In~\cite{Zeng18}, convex optimization and linear programming were utilized to find the optimal set of intermediate points and speed for a UAV. The UAV trajectory is optimized to minimize the connection time constraint with the ground BSs.
	Graph theory and convex optimization have also been used to ensure that the UAV remains connected to at least one BS while minimizing travel time~\cite{Zhang19UAV}.
	However, the above-mentioned conventional optimization-based UAV trajectory design methods often face practical limitations, as many of the optimization problems are non-convex and difficult to solve effectively.
	
	Alternatively, machine learning (ML) techniques have emerged as a promising alternative for UAV trajectory design~\cite{Zeng2017energy, Zhang2020radio, Challita18, Challita2019machine}. 
	For instance, the Q-learning method has been used to design trajectories that maximize continuous connection time~\cite{Khamidehi20}, while a dueling double Q network has been employed to optimize UAV trajectory~\cite{Zeng20}. 
	Additionally, deep reinforcement learning (DRL) has also shown great promise in addressing complex decision-making problems associated with UAV trajectory design.
	For example, a DRL algorithm, based on echo state network cells, was developed for UAV path planning~\cite {Challita19}, and DRL was also utilized to enhance the ability of UAVs to understand the environment based on the captured image~\cite{Arafat23}. However, one of the main challenges associated with these ML techniques is the need for extensive measurement campaigns, which can be prohibitively expensive for large-scale scenarios or complex environments. 
	Deterministic models such as ray tracing have been utilized to generate radio maps for UAV trajectory design, but these methods do not take into account factors such as rain attenuation, which can significantly affect the performance of cellular-connected UAVs.
	%	With the widespread use of high-frequency wireless communication systems, it is important to consider the impact of dielectric attenuation caused by rain in atmospheric environments during trajectory design for UAVs. 
	%Trajectory design for cellular-connected UAV in a given airspace with rain medium has not been adequately explored yet.
	
	Rain attenuation is a well-studied phenomenon in radio wave propagation, and several empirical and semi-empirical models have been proposed to predict the attenuation caused by rain~\cite{Abdulrahman2011empirically, Abdulrahman2012rain}. 
	However, it should be noted that these models are primarily focused on point-to-point or point-to-multipoint attenuation, and do not account for the effects of multipath propagation. Therefore, there is a need for further research to explore and develop techniques that can accurately model the impact of multipath propagation on rain attenuation prediction. The parabolic wave equation (PWE) method has shown promise in modeling wave propagation, as it can properly take into account wave refraction and diffraction~\cite{Levy00, Sheng2016study, Sheng2013modeling}. Additionally, the PWE method has proved to be effective in predicting radio wave propagation under complex meteorological conditions~\cite{He18Wave}.
	
	\subsection{Contributions}
	Motivated by the above facts, we propose a deep reinforcement learning-based framework, which can design trajectories for cellular-connected UAVs in rainy environments using rainfall parameters and detailed information about the environment's geometry. 
	The main contributions of this paper are summarized as follows:
	\begin{itemize}
		\item Firstly, we formulate the UAV trajectory optimization problem to minimize the difference between the mission completion time and the weighted signal-to-interference ratio (SIR). We then transform this problem into a sequential decision-making one, and reformulated it as a Markov decision process (MDP) with a designed state space, action space, and reward function.
		\item Secondly, we employ a physics-based deterministic wave propagation model to extract the SIR, replacing the time-consuming measurement process. This model allows us to consider the detailed information about the environment's geometry and the impact of meteorological conditions on radio wave propagation and trajectory design.
		\item Thirdly, we utilize deep reinforcement learning to address the reformulated MDP problem, and this technique can learn policies in MDP. Besides, deep neural networks (DNN) are used to approximate the Q-function in DRL. We employ a dueling network architecture with a double deep Q-network (DDQN) to train the DNN. This architecture uses separate sets of weights to select actions and evaluate their Q-values, which reduces overestimation and improves stability.
		\item Finally, we conduct extensive evaluations of our proposed approach and investigate the impact of various rainy weather conditions on UAV trajectory design. We compare the optimal trajectory for cellular-connected UAVs in the atmosphere medium and rain medium, highlighting the necessity of accurate physics-based wave propagation models for UAV trajectory design. We also present a comparative analysis of different rainfall parameters and weight coefficients in the problem. Our results demonstrate the effectiveness of our approach in managing varying weather conditions, and we show that weight coefficients can be adjusted based on the priorities of the users, whether they are more concerned with flying time or SIR.
	\end{itemize}
	
	The rest of this paper is organized as follows.
	In Section \ref{Section: model_problem}, we present the system model and formulate the trajectory design problem. 
	Section \ref{Section: rain} provides an introduction to the wave propagation model in rainy environments, while Section \ref{Section: preliminaries} gives an overview of deep Q learning.
	In Section \ref{Section: proposed}, we present our proposed DRL-based trajectory design, while the numerical results and performance evaluation of our approach are given in Section \ref{Section: results}.
	Finally, we conclude the paper in Section \ref{Section: conclusion}.

	%%%%%%%%%%%%%%%%%%%%%%%%%%%%%%%%%%%%%%%%%%%%%%%%%%%%%%%%%%%%%%%%%%%%%%%%%%%%%%%%
	%%%%%%%%%%%%%%%%%%%%%%%%%%%%%%%%%%%%%%%%%%%%%%%%%%%%%%%%%%%%%%%%%%%%%%%%%%%%%%%%
	\section{System Model and Problem Formulation}
	\label{Section: model_problem}
\begin{figure}[!htb]\centering
		\vspace*{-0.2cm}
		\begin{tikzpicture}
			\node[] at(0,0){\includegraphics[scale=0.26,clip,trim={0cm 0cm 0cm 0cm}]{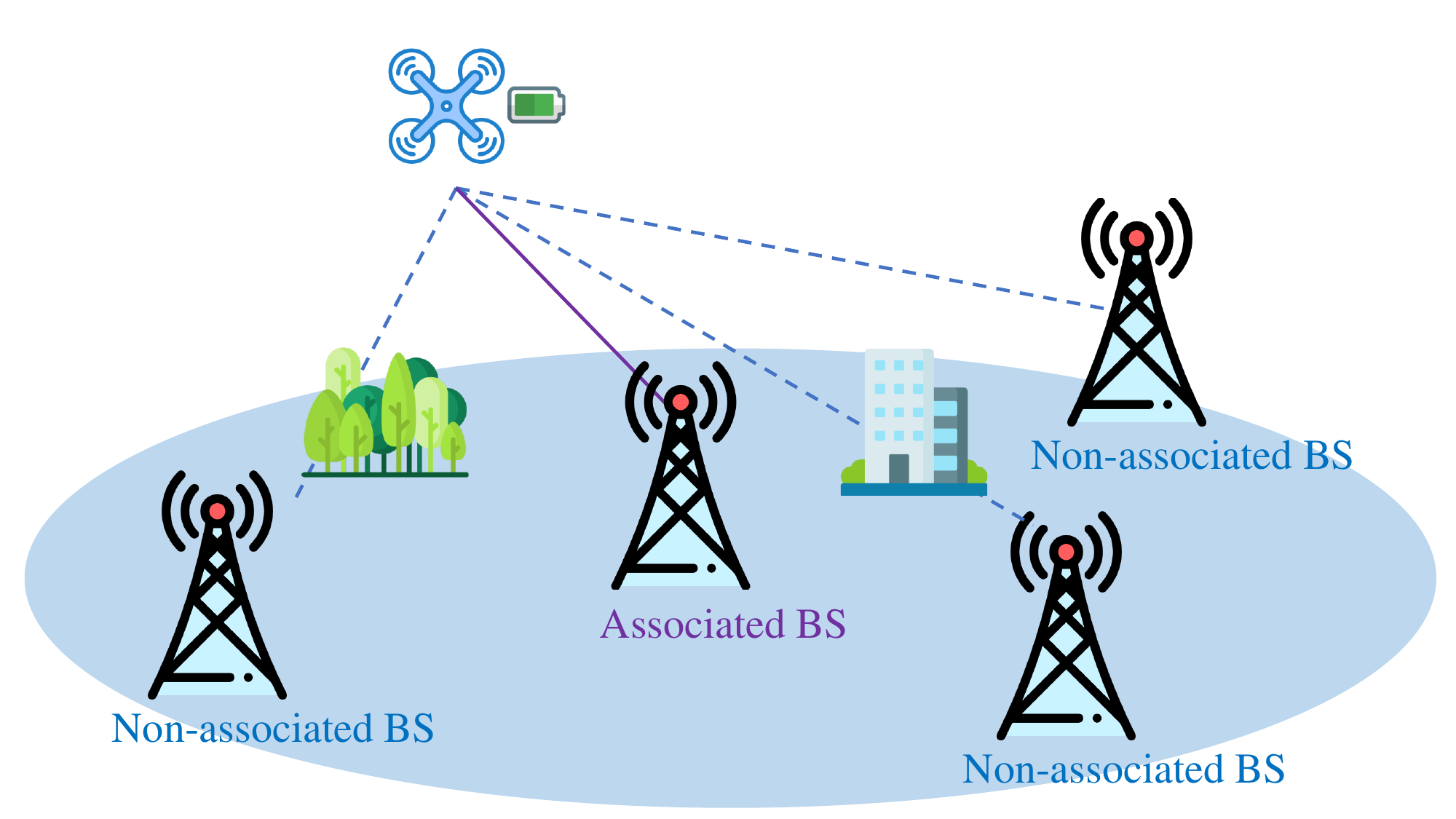}};
		\end{tikzpicture}
		\vspace*{-0.2cm}
		\caption{Cellular-connected UAV communication systems.}
		\label{fig: UAV_scenario}
	\end{figure}

	%\subsection{System Model}
	As shown in~\figref{fig: UAV_scenario}, a cellular-connected UAV system is considered, where UAV is supported by cellular BSs.
	The considered airspace can be represented as a cubic volume, which can be specified by $\textit{X}\times\textit{Y}\times\textit{Z}$, where $\textit{X}\in[x_L,x_H]$, $\textit{Y}\in[y_L,y_H]$, and $\textit{Z}\in[z_L,z_H]$. 
	
\begin{figure}[!htb]\centering
		\vspace*{-0.1cm}
		\begin{tikzpicture}
			\node[] at(0,0){\includegraphics[scale=0.26,clip,trim={0cm 0cm 0cm 0cm}]{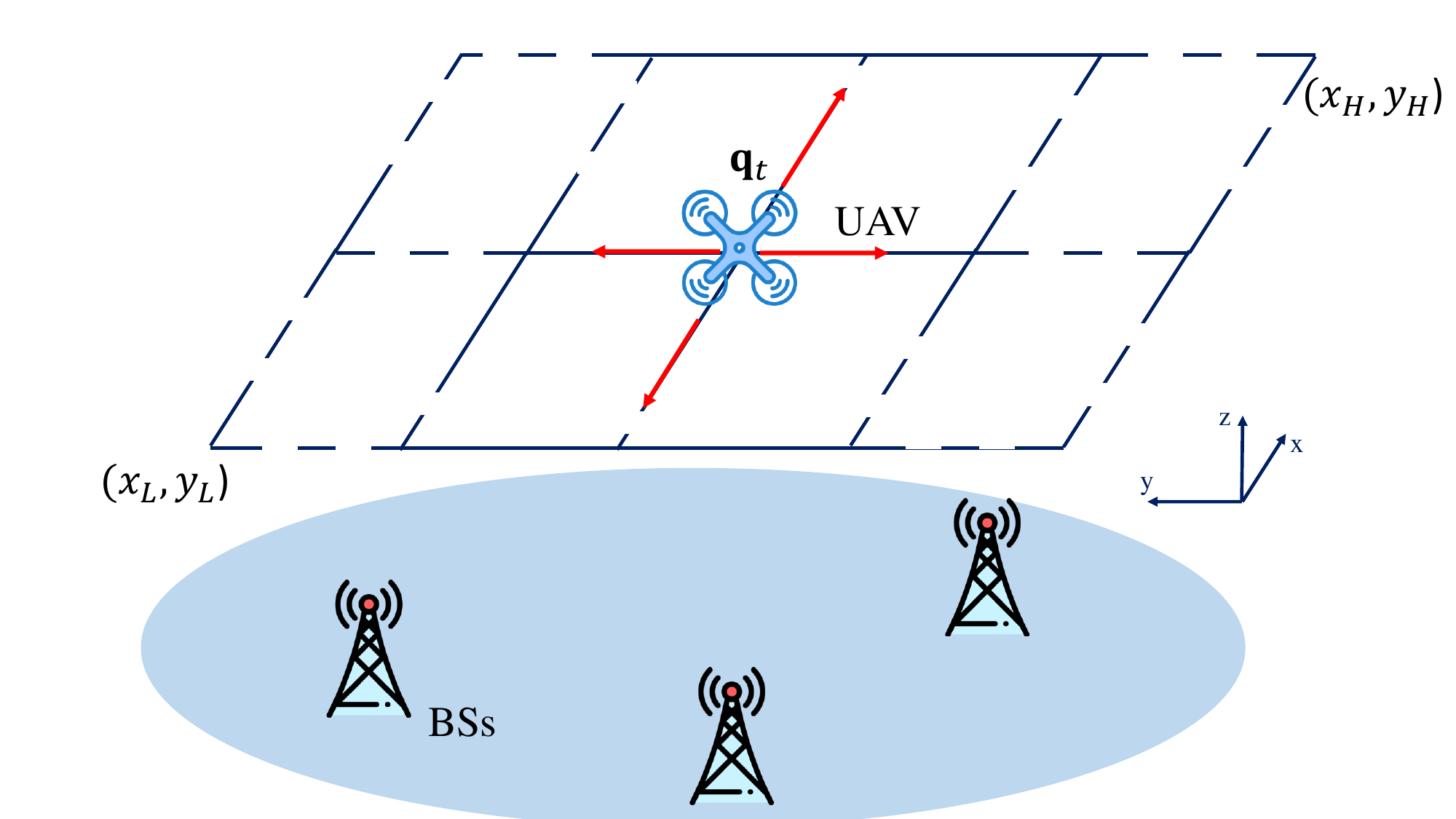}};
		\end{tikzpicture}
		\vspace*{-0.1cm}
		\caption{Illustration of available grid points in the airspace.}
		\label{fig: UAV_grid}
	\end{figure}

	We assume that the UAV takes off from a random initial location, $\mathbf{q}_{I}$, and flies towards a fixed final destination, $\mathbf{q}_{F}$.
	If a fixed height is assumed for the UAV's flight, we can visualize the range of movement for the UAV as \figref{fig: UAV_grid}, where $(x_L,y_L)$ and $(x_H,y_H)$ are the boundary points of the considered airspace at the fixed height.
	An optimal trajectory can be designed for the UAV to minimize its flying time while maintaining reliable communication connectivity with BSs.
	The UAV trajectory can be represented by $\mathbf{q}(t), t\in [0, T]$, where $T$ denotes the flying time. 
	Hence, we have:
	\begin{equation}
		\mathbf{q}(0) = \mathbf{q}_{I},
	\end{equation}
	\begin{equation}
		\mathbf{q}(T) = \mathbf{q}_{F},
	\end{equation}
	\begin{equation}
		\mathbf{q}_L \preccurlyeq \mathbf{q}(t) \preccurlyeq \mathbf{q}_H, \forall t \in (0,T).
	\end{equation}
	where $\preccurlyeq$ indicates the element-wise inequality.
	The coordinates of $\mathbf{q}_L$ and $\mathbf{q}_H$ are $(x_L,y_L,z_L)$ and $(x_H,y_H,z_H)$, respectively.
	
	In particular, the whole considered airspace is divided into $M$ cells, and $u_m(t)$, $1\leq m\leq M$, represents the received signal strength (RSS) from cell $m$ to the UAV at time $t$. RSS in the rain medium can be generated by the PWE simulator, and a detailed description of the PWE method is provided in Section \ref{Section: rain}. 
	As the UAV is associated with one BS at a time, we define $b(t) \in \{1,..., M\}$ to denote the cell which is associated with the UAV at time $t$.
	Then, SIR at time $t$ can be defined as:
	\begin{equation}
		\text{SIR}\left(\mathbf{q}(t), b(t)\right) = \frac{[u_{b(t)}(t)]^2}{\sum_{m \neq b(t)}^{}[u_{m}(t)]^2}.
		\label{eq: SIR}
	\end{equation}
	The total expected SIR along the UAV path can be expressed as:
	\begin{equation}
		\text{SIR}_{\text{total}}\left(\mathbf{q}(t), b(t)\right) = \int_{0}^{T} \text{SIR}\left(\mathbf{q}(t), b(t)\right) dt.
	\end{equation}
	
	Intuitively, the UAVs can easily avoid weak coverage regions of the cellular network in the sky with longer flying times $T$.
	However, practical applications require reducing the flying time due to the limited endurance of UAVs.
	Hence, there should be a tradeoff between minimizing the UAV flying time and maximizing the total SIR along the trajectory, which can be balanced by designing $\{\mathbf{q}(t)\}$ and $\{b(t)\}$.
	The trajectory design problem can be formulated as follows by introducing a certain weight $\mu$:
	\begin{equation}
		(\text{P1}):   \\
			\max \limits_{T, \{\mathbf{q}(t),b(t), \overrightarrow{\mathbf{v}}(t)\}}  -T+\mu \text{SIR}_{\text{total}}\left(\mathbf{q}(t), b(t)\right)\\ 
		\label{eq:P1}
	\end{equation}
	\begin{align}
		\re{\text{s.t.}} & \re{\left \|  \mathbf{q}(t+1)-\mathbf{q}(t) \right\| \leq V_{\text{max}}\overrightarrow{\mathbf{v}}(t),\forall t \in [0,T-1], }\tag{\ref{eq:P1}a}	\\
		&\re{\mathbf{q}(0) = \mathbf{q}_{I},}\tag{\ref{eq:P1}b}	\\
		& \re{\mathbf{q}(T) = \mathbf{q}_{F}, }\tag{\ref{eq:P1}c}\\
		& \re{\mathbf{q}_L \preccurlyeq \mathbf{q}(t) \preccurlyeq \mathbf{q}_H, \forall t \in [0,T],}\tag{\ref{eq:P1}d}\\
		& \re{b(t) \in \{1,...,M\},}\tag{\ref{eq:P1}e}\\
		& \re{\left \|\overrightarrow{\mathbf{v}}(t)\right\| = 1, \forall t \in [0,T],} \tag{\ref{eq:P1}f} \\
		& \re{\text{SIR}_\text{min} \leq \text{SIR}\left(\mathbf{q}(t), b(t)\right),} \tag{\ref{eq:P1}g}
	\end{align}
	where $V_\text{max}$ denotes the maximum UAV speed and  $\overrightarrow{\mathbf{v}}(t)$ is the UAV flying direction.
	\re{$\text{SIR}_\text{min}$ indicates the minimum SIR along the UAV path. Its specific definition can be tailored to suit the requirements of the actual applications being considered.}
	Note that the weight $\mu$ is a non-negative coefficient.
	A higher value of the parameter $\mu$ indicates that the algorithm prioritizes cellular connectivity with the UAV over the flying time.
	
	However, in practice, the optimization problem (P1) is highly non-convex, making it challenging to be efficiently solved.
	In addition, obtaining SIR in rain medium through measurement campaigns is a time-consuming and labor-intensive process. 
	To address these issues, we propose a solution that employs a physics-based wave propagation simulator, PWE, along with a DRL algorithm based on a dueling deep network with multi-step learning.

	%\subsection{Problem Formulation}

	%\vspace{0.5cm}	
	%
	%%%%%%%%%%%%%%%%%%%%%%%%%%%%%%%%%%%%%%%%%%%%%%%%%%%%%%%%%%%%%%%%%%%%%%%%%%%%%%%%
	%%%%%%%%%%%%%%%%%%%%%%%%%%%%%%%%%%%%%%%%%%%%%%%%%%%%%%%%%%%%%%%%%%%%%%%%%%%%%%%%
	\section{Physics-Based Wave Propagation Model in Rain Medium}
	\label{Section: rain}
	The parabolic wave equation method~\cite{Levy00} is a physics-based wave propagation modeling approach and can be expressed as
	\begin{equation}
		\frac{\partial{u}}{\partial{z}}=\frac{1}{2\text{j}k_0}\left(\frac{\partial^2}{\partial{x}^2}+\frac{\partial^2}{\partial{y}^2}\right)u.
		\label{eq: PE}
	\end{equation}
	In \eqref{eq: PE}, $u$ is the reduced plane wave solution, and radio waves are assumed to propagate along the $z$-axis. 
	\re{In general, \eqref{eq: PE} can be solved using different techniques, such as the finite difference method and the split-step Fourier method. It is found that the split-step Fourier scheme is more time-efficient compared to the finite difference scheme with the frequency increases, which makes it a good candidate for high-frequency propagation modeling~\cite{QIN21efficient}.
	Hence, in the paper, \eqref{eq: PE} is solved using a split-step Fourier technique as }
	\begin{equation}
		\begin{split}
			u(x,y,z+\varDelta z)= e^{-\text{j}k_0\left(n^2 - 1\right)\varDelta x/2}\times \\
			\mathcal{F}^{-1}\left\{C\left(k_x \right)C\left(k_y \right) \mathcal{F}\left\{u\left(x,y,z\right)\right\} \right\},
		\end{split}
		\label{eq: SSPE}		
	\end{equation}
	where $\mathcal{F}{}$ and $\mathcal{F}^{-1}$ are the Fourier transform pairs; $k_x$ and $k_y$ are the spectral variables, and 
	\begin{equation}
		%		C\left(k\right)=exp(\frac{-jk^2\varDelta z}{k_0+\sqrt{{k_0}^2 - k^2}})
		C\left(k\right)=\exp(\frac{-\text{j}k^2\varDelta z}{2k_0}).
		\label{eq: 00}		
	\end{equation}
	In \eqref{eq: SSPE}, $n=\sqrt{\varepsilon_{\text{eff}}}$ is the complex refractivity that can be calculated accordingly for the medium, and $\varepsilon_{\text{eff}}$ is the relative effective permittivity of the medium.
	
	The effective permittivity $\varepsilon_{\text{eff}}$ can be defined as
	\begin{equation}
		\varepsilon_{\text{eff}} = \varepsilon_{0}+\frac{\bar{P}}{\bar{E}},
		\label{eq: eff}
	\end{equation}
	where $\varepsilon_{0}$ is the permittivity of the air, $\bar{E}$ is the field in the medium, and $\bar{P}$ denotes the average polarization density of the raindrops.
	In particular, $\bar{P}$ can be calculated based on the raindrop spectrum $N(D)$, the polarization rate $\alpha$, and the field $\bar{E}_e$ inside the raindrops~\cite{Gong2006accurate}:
	\begin{equation}
		\bar{P} = \int_{D_\text{min}}^{D_\text{max}} N(D)\alpha(D)\bar{E}_e(D) dD,
		\label{eq: P}
	\end{equation}
	where $D$ is the equivalent diameter of a single raindrop, and
	$\bar{E}_e(D)$ is the field inside raindrop. If the raindrop is a sphere:
	\begin{equation}
		\bar{E}_e = \bar{E} + \frac{\bar{P}}{3\varepsilon_{0}}.
	\end{equation}
	If raindrop is an ellipsoid:
	\begin{equation}
		\bar{E}_e = \bar{E} + \frac{L_i\bar{P}}{\varepsilon_{0}}.
	\end{equation}
	$L_i$ denotes the polarization factor, which can be expressed as:
	\begin{equation}
		L_a = \frac{1}{e^2}\left(1-\sqrt{\frac{1-e^2}{e^2}}\arcsin e\right),
	\end{equation}
	\begin{equation}
		L_b = L_c = \frac{1}{2} - \frac{1}{2}L_a,
	\end{equation}
	where $e$ is the eccentricity of an ellipsoidal raindrop that can be found in 
	\begin{equation}
		e = \sqrt{1-\left(\frac{a}{b}\right)^2},
	\end{equation}
	
	The rain medium is generally formed by raindrops of various shapes and sizes in the atmosphere. Since the drops with a diameter larger than 8\,mm are unstable and can easily break up~\cite{Sheng2013modeling}, we assume that the range for the diameter of raindrops is from 0.1\,mm to 8\,mm. 
	If the diameter of the raindrop is more than 1.25\,mm, the shape of the raindrop is a flat rotary ellipsoid. Otherwise, the shape can be assumed to be spherical. For the polarization rate $\alpha$, if raindrop is a sphere,
	\begin{equation}
		\alpha(D) = 4\pi \frac{\varepsilon_{0}(\varepsilon_{\omega}-\varepsilon_{0})}{\varepsilon_{\omega}-2\varepsilon_{0}}(\frac{D}{2})^3.
	\end{equation}
	If raindrop is an ellipsoid,
	\begin{equation}
		\alpha(D) = v \frac{\varepsilon_{0}(\varepsilon_{\omega}-\varepsilon_{0})}{\varepsilon_{\omega}+L_i(\varepsilon_{\omega}-\varepsilon_{0})},
		\label{eq: alpha}
	\end{equation}
	where $v$ is the volume of one raindrop, and $\varepsilon_{\omega}$ is the permittivity of water, which can be calculated by the Debye formula~\cite{Liebe1989millimeter}.
	Substituting \eqref{eq: P} - \eqref{eq: alpha} into \eqref{eq: eff}, we can obtain the effective permittivity in rain medium:
	\begin{equation}
		\begin{split}
			&\varepsilon_{\text{eff}} = \varepsilon_{0} + \int_{D_\text{min}}^{1.25} N(D)4\pi \frac{\varepsilon_{0}(\varepsilon_{\omega}-\varepsilon_{0})}{\varepsilon_{\omega}-2\varepsilon_{0}}(\frac{D}{2})^3\\
			&\times\frac{3\varepsilon_{0}}{3\varepsilon_{0}-4\pi \displaystyle{\frac{\varepsilon_{0}(\varepsilon_{\omega}-\varepsilon_{0})}{\varepsilon_{\omega}-2\varepsilon_{0}}}(\frac{D}{2})^3} dD\\
			&+ \frac{1}{3}\sum_{i=1}^{3}\int_{1.25}^{D_\text{max}} N(D)\frac{4\pi}{3}\frac{\varepsilon_{0}(\varepsilon_{\omega}-\varepsilon_{0})}{\varepsilon_{\omega}+L_i(\varepsilon_{\omega}-\varepsilon_{0})} (\frac{D}{2})^3 \\
			&\times \frac{\varepsilon_{0}}{\varepsilon_{0}-L_i \displaystyle{\frac{4\pi}{3}\frac{\varepsilon_{0}(\varepsilon_{\omega}-\varepsilon_{0})}{\varepsilon_{\omega}+L_i(\varepsilon_{\omega}-\varepsilon_{0})}} (\frac{D}{2})^3} dD.
		\end{split}
	\label{eq: eff_rain}
	\end{equation}
	where the raindrop spectrum is chosen as the Marshall-Palmer spectrum, which is a function of the rain rate $R$ (mm/h).
	
\begin{figure}[!htb]\centering
		\vspace*{-0.8cm}
		\begin{tikzpicture}
			\node[] at(0,0){\includegraphics[scale=0.5,clip,trim={0cm 0cm 0cm 0cm}]{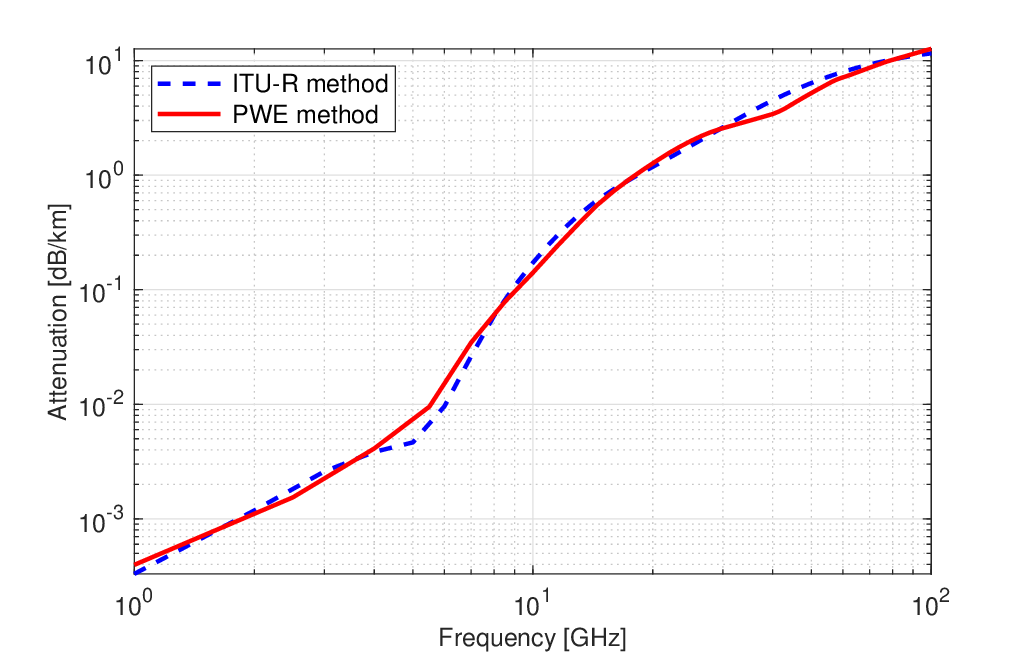}};
		\end{tikzpicture}
		\vspace*{-0.8cm}
		\caption{Rain attenuation obtained from the PWE simulator and the ITU-R model at different frequencies.}
		\label{fig: Attenuation_frequency_rain}
	\end{figure}

	For validation purposes, as shown in \figref{fig: Attenuation_frequency_rain}, we compare the numerical results obtained from the PWE-based electromagnetic simulator at different frequencies against the ITU-R model~\cite{ITU05}, which is a widely used statistical model for predicting the effects of rain on radio wave propagation.
	The attenuation of the ITU-R model depends on the rain rate $R$:
	\begin{equation}
		\gamma_R = k R^\alpha,
	\end{equation}
	where the coefficients $k$ and $\alpha$ can be found in \cite{ITU05}.
	
	In this study, we set the rain rate as $R = 12.5$\,mm/h and compared the attenuation results obtained from the PWE simulator and the ITU-R model at different frequencies, as shown in \figref{fig: Attenuation_frequency_rain}. The results demonstrate a good match between the PWE simulator and the ITU model, indicating the reliability of using the PWE simulator in rainy environments.
	
	It's worth noting that the ITU-R model relies on a large database of experimental measurements to provide empirical relationships, but it has limitations in complex environments where terrain and obstacles can significantly affect wave propagation. Additionally, the ITU-R model cannot provide a 3D received signal strength distribution in a given skyspace. Therefore, in our proposed approach, we employed the PWE simulator to obtain the electromagnetic wave propagation characteristics in rainy environments.
	
	%%%%%%%%%%%%%%%%%%%%%%%%%%%%%%%%%%%%%%%%%%%%%%%%%%%%%%%%%%%%%%%%%%%%%%%%%%%%%%%%
	%%%%%%%%%%%%%%%%%%%%%%%%%%%%%%%%%%%%%%%%%%%%%%%%%%%%%%%%%%%%%%%%%%%%%%%%%%%%%%%%
	\section{Preliminaries: Deep Q Network}
	\label{Section: preliminaries}
	In this section, we provide a brief overview of the deep Q network (DQN). A more comprehensive description can be found in~\cite{Sutton18}.
	DQN is a widely adopted reinforcement learning method designed to solve MDP in continuous state spaces. In DQN, an agent interacts with the environment to collect data to train itself.
	Unlike basic reinforcement learning in tabular settings, which stores the expected return of each action under a specific state in tabular form, DQN parameterizes the expected return of each action under a specific state using deep neural networks (DNNs).
	As a result, the policy of the DQN agent is governed by the parameters of DNN, which are updated using gradient descent with respect to the expected returns calculated using \emph{bootstrapping}\cite{Sutton18}.
	
	Specifically, at the start of the training, a DQN network with random coefficients $\pmb{\theta}$ is initialized.
	This network is referred to as the Q network.
	At each state $s \in \mathcal{S}$, the Q network selects the action that maximizes its output, known as the \emph{exploitation} of the known expected returns predicted by the Q network.
	Additionally, with a probability of $\epsilon$, the Q network selects a random action from the available actions, known as the \emph{exploration}. This allows for unexplored action-state pairs to be updated in the current Q network. The parameter $\epsilon$ is used to balance the exploitation and exploration and decreases as a function of training episodes. This policy for selecting an action is known as the $\epsilon$-greedy policy and is shown as follows:
	\begin{equation}
		\label{eq: e-greedy}
		f(a,b) = 
		\begin{cases}
			a \leftarrow a_{id}, id \sim \mathcal{U}\{1,|\mathcal{A}|\} 
			\\ \hspace{2.45cm}\textrm{with probability $\epsilon$},\\
			a \leftarrow \underset{a}{\textrm{argmax}} \hat{Q}(s,a;\pmb{\theta}), a_{id}\in\mathcal{A} 
			\\\hspace{1.9cm} \textrm{with probability $1-\epsilon$}.\\
		\end{cases}
	\end{equation}
	
	Once an action is selected according to \eqref{eq: e-greedy}, the next state $s_{next}$ can be obtained using current state $s$, action $a$ and transition model $P(s_{next}|s, a)$. In our settings, the transition is deterministic, specifically, $s_{next} = s + a$  with $P(s_{next}|s, a)=1$. Before moving to the next state, the reward $R$ at the current is calculated. The $(s, R, a, s_{next})$ tuple is the data that the DQN agent collected in the training phase for the training of the DNN. Such a tuple is known as a \emph{transition}. With such transitions, expected returns of action $a$ at state $s$ can be estimated using bootstrapping and Monte Carlo (MC) methods. Specifically, transitions are stored in a replay memory $D$ with capacity $C$. Then, a minibatch of $B$ transitions is randomly sampled from $D$ with $B\ll C$. The expected returns of the sample transitions are estimated using bootstrapping in \eqref{eq: bootstrapping}, where $\gamma$ is the discount factor. \eqref{eq: bootstrapping} uses the next action that maximizes the expected return predicted by the Q network state $s_{next}$ instead of the \emph{actual action} taken at $s_{next}$, which is a method used in Q learning to allow quick convergence.
	\begin{equation}
		\label{eq: bootstrapping}
		\mathbb{E}(Q_{s,a}) = 
		\begin{cases}
			R & \textrm{if episode ends},\\
			R + \gamma\underset{a\in\mathcal{A}}{\textrm{max}}\hat{Q}(s_{next},a;\pmb{\theta}) & \textrm{else}.\\
		\end{cases}
	\end{equation}
	
	The DNN parameters $\pmb{\theta}$ can be trained using gradient descent with respect to the expected returns. The loss function is formulated in \eqref{eq: criticLoss}:
	\begin{equation}
		\label{eq: criticLoss}
		L(\pmb{\theta}) = \lVert \hat{Q}(s,a;\pmb{\theta}) - \mathbb{E}(Q_{s,a})) \rVert.
	\end{equation}
	However, $\mathbb{E}(Q_{s, a})$ in \eqref{eq: bootstrapping} is known to be a noisy estimation with high variance. This is because, in \eqref{eq: bootstrapping}, the estimation itself depends on the parameters that need to be trained.
	A few techniques are used in addition to the above DQL to further stabilize and speed up the training process. These include the \emph{target network}\cite{Mnih15} and the \emph{multi-step learning}\cite{Hessel18}. Target network suggests the maintenance of a \emph{target network} which is updated after multiple episodes. The target network shares the same architecture as the DNN. It is used for the prediction of the expected returns for different actions at $s_{next}$. With a target network, the estimation of the expected returns in \eqref{eq: bootstrapping} becomes:
	\begin{equation}
		\label{eq: bootstrapping_target_network}
		\mathbb{E}(Q_{s,a}) = 
		\begin{cases}
			R & \textrm{if episode ends},\\
			R + \gamma\underset{a\in\mathcal{A}}{\textrm{max}}\hat{Q}(s_{next},a;\pmb{\theta}^{-}) & \textrm{else}.\\
		\end{cases}
	\end{equation}
	
	In \eqref{eq: bootstrapping_target_network}, $\mathbb{E}(Q_{s, a})$ becomes independent of the DNN parameters $\pmb{\theta}$ and the training process becomes more stable. The parameters $\pmb{\theta}^{-}$ of the target network are set to be updated after a few training episodes with $\pmb{\theta^{-}}\leftarrow\pmb{\theta}$.\par
	
	Multi-step learning provides more efficient use of the replay memory and a more stable training process by learning from multi-step bootstrap targets using a sliding window $W$ (a queue) with length $N_1$. $N_1$ transitions are stored in $W$ before being stored in the replay memory. The transitions sliding window are used to calculate the \emph{multi-step reward}. Specifically, for state-action pair $(s,a)$, its multi-step reward $R_{(0:N_1)}$ is calculated using \eqref{eq: NStepReward} with $n=0$ ($(s,a)$ as the current state).
	\begin{equation}
		\label{eq: NStepReward}
		R_{n:(n+N_1)} = \sum^{N_1-1}_{i=0}\gamma^{i}R_{n+i+1}.
	\end{equation}
	Then $(s_n, a_n, R_{(0:N_1)}, s_{n+N_1})$ is stored as a transition into the replay memory $D$. After the calculation of $R_{(0:N_1)}$, $(s, a)$ is popped out of the head of $W$, and a new transition is appended to the end of $W$. Now the calculation of $\mathbb{E}(Q_{s,a})$ becomes:
	\begin{equation}
		\label{eq: bootstrapping_target_network_nStep}
		\mathbb{E}(Q_{s,a}) = 
		\begin{cases}
			R_0 & \textrm{if episode ends, }\\
			R_{(0:N_1)} + \gamma\underset{a\in\mathcal{A}}{\textrm{max}}\hat{Q}(s_{next},a;\pmb{\theta}^{-}) & \textrm{else}.\\
		\end{cases}
	\end{equation}
	
	In this paper, a dueling DQN~\cite{Wang16} is adopted for the DNN architecture. Compared to the standard DQN, dueling DQN is able to learn the state-value function more efficiently.

	%\vspace{0.3cm}
	%%%%%%%%%%%%%%%%%%%%%%%%%%%%%%%%%%%%%%%%%%%%%%%%%%%%%%%%%%%%%%%%%%%%%%%%%%%%%%%%
	%%%%%%%%%%%%%%%%%%%%%%%%%%%%%%%%%%%%%%%%%%%%%%%%%%%%%%%%%%%%%%%%%%%%%%%%%%%%%%%%
	\section{Proposed Trajectory Design Approach for Cellular-Connected UAVs}
	\label{Section: proposed}
	In this section, we present our proposed DRL-based approach for solving the trajectory design problem of cellular-connected UAVs in rainy environments.
	
	\subsection{Reformulation as an MDP}
	To solve the trajectory design problem (P1) using reinforcement learning methods, the problem can be first reformulated as an MDP, where the future states are only affected by the current state of the system and the selected action of the agent.
	First, the flying time $T$ can be separated into $N$ discrete time steps. 
	We use $\Delta t$ to present the time interval and $T = N\Delta t$.
	Subsequently, the UAV trajectory $\mathbf{q}(t)$ can be represented by a sequence $\{\mathbf{q}_1,...,\mathbf{q}_N\}$.
	Hence, (6a) - (6d) can be rewritten as:
	\begin{align}
		&\left \|  \mathbf{q}_{n+1}-\mathbf{q}_n \right\| \leq V_{\text{max}}\overrightarrow{\mathbf{v}}(t),\forall n,\forall t \in [0,T-1],\\ 
		&\mathbf{q}_0 = \mathbf{q}_{I},\\
		& \mathbf{q}_N = \mathbf{q}_{F},\\
		& \mathbf{q}_L \preccurlyeq \mathbf{q}_n \preccurlyeq \mathbf{q}_H, \forall n.
	\end{align}
	Similarly, the association policy $b(t)$ can be represented as $b_n$. Besides, the UAV flying direction is approximated as $\overrightarrow{\mathbf{v}}_n = \overrightarrow{\mathbf{v}}(n*\Delta t)$ at time $t$.
	As a result, (6e) and (6f) can be written as:
	\begin{align}
		& b_n \in \{1,...,M\},\\
		& \left \|\overrightarrow{\mathbf{v}}_n\right\| = 1, \forall n.
	\end{align}
	
	For each step $n$, SIR in rain medium can be calculated by the PWE simulator introduced in Section \ref{Section: rain}. The location $\mathbf{q}_n$, the association policy $b_n$, and the information of the environment are all taken into account.
	Based on the above discussions, the problem (P1) can be approximated as follows:
	\begin{equation}
			(\text{P2}):\\
		\max \limits_{N, \{\mathbf{q}_n,b_n\overrightarrow{\mathbf{v}}_n\}} -N+\mu \sum_{n=1}^{N} \text{SIR}\left(\mathbf{q}_n, b_n\right)\\
		\label{eq:P2}
	\end{equation}
	\begin{align}
		\re{\text{s.t.}} & \re{\left \|  \mathbf{q}_{n+1}-\mathbf{q}_n \right\| \leq V_{\text{max}}\overrightarrow{\mathbf{v}}_n,\forall n,} \tag{\ref{eq:P2}a}\\ 
		&\re{\mathbf{q}_0 = \mathbf{q}_{I}, }\tag{\ref{eq:P2}b}\\
		& \re{\mathbf{q}_N = \mathbf{q}_{F},} \tag{\ref{eq:P2}c}\\
		& \re{\mathbf{q}_L \preccurlyeq \mathbf{q}_n \preccurlyeq \mathbf{q}_H, \forall n,} \tag{\ref{eq:P2}d}\\
		& \re{b_n \in \{1,...,M\}, }\tag{\ref{eq:P2}e}\\
		& \re{\left \|\overrightarrow{\mathbf{v}}_n\right\| = 1, \forall n,} \tag{\ref{eq:P2}f}\\
		& \re{\text{SIR}_\text{min} \leq \text{SIR}_n, \forall n.} \tag{\ref{eq:P2}g}
	\end{align}
	Furthermore, the MDP can be formulated as follows:
	\subsubsection{States}
	The state space is constructed of the potential UAV location, $\mathbf{q}_n$, in the considered airspace.
	It is defined as 
	\begin{equation}
		\mathcal{S} = \{\mathbf{q}:\mathbf{q}_L\preccurlyeq \mathbf{q} \preccurlyeq \mathbf{q}_U\}.
	\end{equation}
	
	\subsubsection{Actions}
	The action space corresponds to the UAV's flying direction. For each time step, the UAV carries out an action. 
	The action space can be defined as
	\begin{equation}
		\mathcal{A} = \left\{\overrightarrow{\mathbf{v}} : \|\overrightarrow{\mathbf{v}}_n\right\| = 1\}.
	\end{equation}
	
	\subsubsection{Transition Probabilities}
	The transition probabilities in this problem are deterministic and governed by
	\begin{equation}
		\left \|  \mathbf{q}_{t+1}-\mathbf{q}_t \right\| \leq V_{\text{max}}\overrightarrow{\mathbf{v}}_n,\forall n.
		\label{eq: Probabilites}
	\end{equation}

	\subsubsection{Rewards}
	The reward function consists of two parts, the flying time and the accumulated SIR along the path. The flying time is defined as the time taken for the UAV to move from its current position to the next position after taking action $a$. 
	The SIR  component of the reward is defined to encourage the UAV to avoid weak coverage regions. Specifically, if the UAV enters a location with a small SIR, a penalty with weight $\mu$ is incurred. The SIR penalty component of the reward is given by $\mu \text{SIR}(\mathbf{q}, b)$.
	The reward is defined as follows:
	\begin{equation}
		R(\mathbf{q},b) = -1+\mu \text{SIR}(\mathbf{q}, b).
	\end{equation}
	
	After being formulated as the above MDP, the problem can be solved by applying a dueling DDQN with multi-step learning.
	\re{By utilizing the dueling DDQN architecture, we mitigate the risk of suboptimal policies caused by overestimation bias in Q-value estimates, resulting in more stable learning. Moreover, the integration of double Q-learning further refines value estimation, leading to a more robust action selection. The adoption of multi-step learning boosts efficiency and sample utilization, accelerating the learning process and enabling the UAV to discover better navigation strategies. }
	
	\subsection{Dueling DDQN Learning for Trajectory Design}
\begin{figure}[!htb]\centering
		%\vspace*{-0.1cm}
		\begin{tikzpicture}
			\node[] at(0,0){\includegraphics[scale=0.42,clip,trim={8cm 3cm 5cm 3.2cm}]{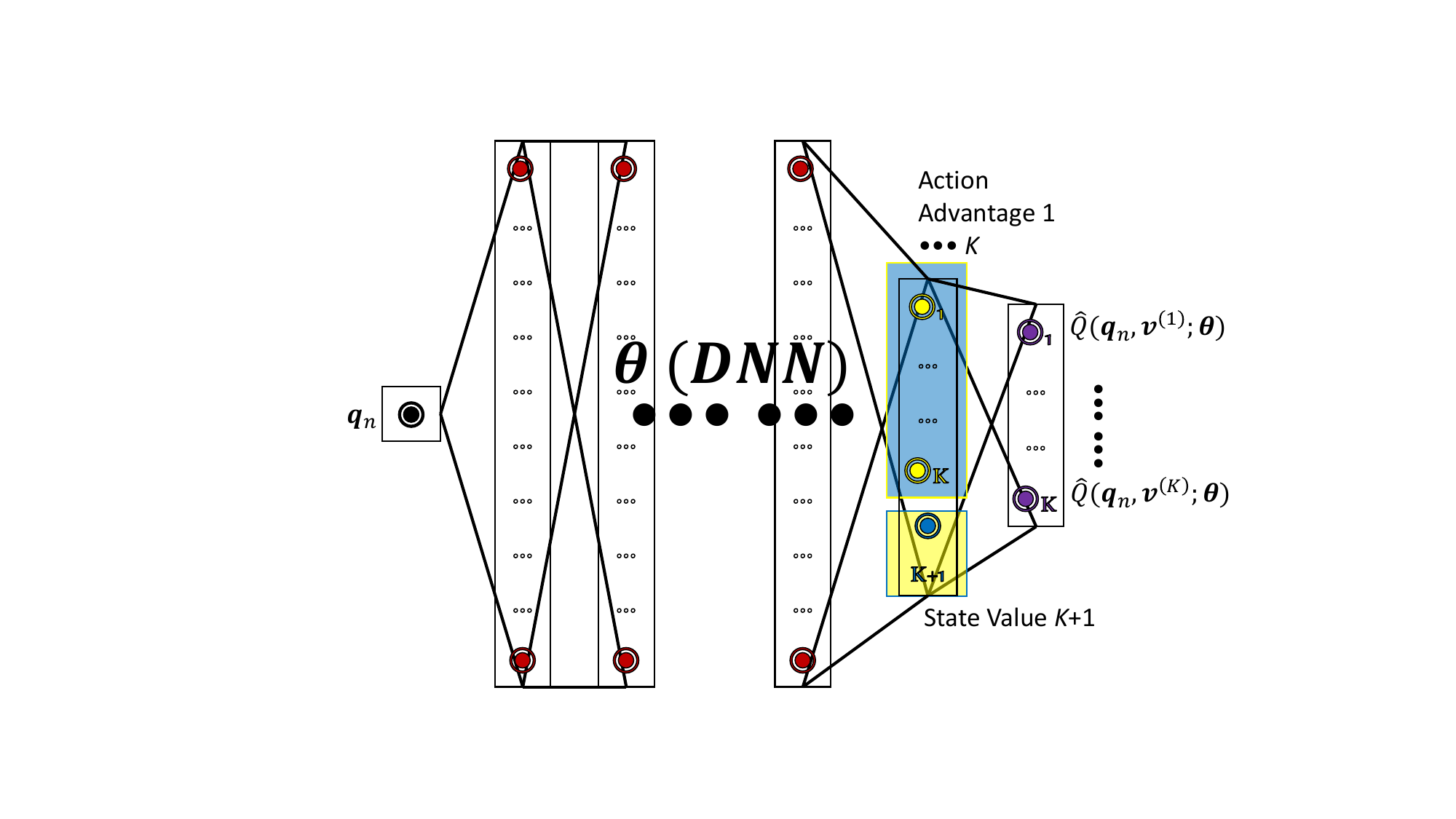}};
		\end{tikzpicture}
		\vspace*{-0.2cm}
		\caption{Diagram of the dueling DQN for the trajectory design problem.}
		\label{fig: Dueling_DQN}
	\end{figure}

	In our proposed approach, we focus on the scenario in which the UAV flies in a horizontal plane with a fixed altitude.
	We use DNN to approximate the action-value function $Q(\mathbf{q},\overrightarrow{\mathbf{v}})$.
	Such a function contains the state input and the action input.
	We discretize the action space $\mathbf{A}$ into $K=4$ values, and $	\mathbf{A} = \{\overrightarrow{\mathbf{v}^{(1)}} ,..., \overrightarrow{\mathbf{v}^{(K)}}\}$.
	The state space $\mathbf{S}$ is continuous. 
	In this paper, the action-value function $Q(\mathbf{q},\overrightarrow{\mathbf{v}})$ is approximated using the dueling DQN, as illustrated in \figref{fig: Dueling_DQN}, where 
	$\theta$ denotes the coefficients of the dueling DQN to ensure that the output $\hat{Q} (\mathbf{q},\overrightarrow{\mathbf{v}};\theta)$ gives a good approximation to the true action-value function $Q(\mathbf{q},\overrightarrow{\mathbf{v}})$.
	At each time step $n$, the input of such network is the location of the UAV, $\mathbf{q}_n$. $K$ outputs correspond to different actions in the action space,  $\mathbf{A}$.
	The multi-step DQN learning is to minimize the loss in \eqref{eq: loss}:
	\begin{equation}
		(R_{n:(n+N_1)} +\gamma^{N_1}\max \limits_{{\mathbf{v}}^{'} \in \mathbf{A}}\hat{Q}(\mathbf{q}_{n+N_1},\overrightarrow{\mathbf{v}}^{'};\pmb{\theta^{-}})-\hat{Q}(\mathbf{q}_{n},\overrightarrow{\mathbf{v}}_{n};\pmb{\theta}) ) ^2
		\label{eq: loss}
	\end{equation}
	
	For multi-step DDQN learning, a truncated $N_1$-step return in \eqref{eq: return} from a given state $\mathbf{q}_n$ is defined.
	\begin{equation}
		R_{n:(n+N_1)} = \sum^{N_1-1}_{i=0}\gamma^{i}R_{n+i+1}.
		\label{eq: return}
	\end{equation}
	Furthermore, the loss of the multi-step learning in DDQN is
	\begin{equation}
		(R_{n:(n+N_1)} +\gamma^{N_1} \hat{Q}(\mathbf{q}_{n+N_1},\overrightarrow{\mathbf{v}}^{*};\pmb{\theta^{-}})-\hat{Q}(\mathbf{q}_{n},\overrightarrow{\mathbf{v}}_{n};\pmb{\theta}) ) ^2,
		\label{eq: loss_DDQN}
	\end{equation}
	where 
	\begin{equation}
		\overrightarrow{\mathbf{v}}^{*}=\underset{\overrightarrow{\mathbf{v}}^{'}\in\mathcal{A}}{\textrm{agrmax}}\hat{Q}(\mathbf{q}_{j+N_1},\overrightarrow{\mathbf{v}}^{'};\pmb{\theta}).
	\end{equation}
	Coefficients $\theta^{-}$ are used to evaluate the bootstrapping action in \eqref{eq: loss_DDQN}.
	
	\begin{algorithm}
		\fontsize{11}{10}\selectfont
		\caption{\fontsize{11}{10}\selectfont Dueling DDQN Multi-Step Learning for Trajectory Design in Rain Medium}
		\label{alg: DDQNMSRL1}
		\begin{algorithmic}[1]
			\STATE \textbf{Initialize:} maximum number of episodes $\overline{N}_{epi}$, maximum number of steps per episode $\overline{N}_{step}$, update frequency $N_{update}$, reaching-destination toleration distance $D_{tol}$, initial exploration $\epsilon_0$, decaying rate $\alpha$, set $\epsilon\leftarrow\epsilon_0$.
			\STATE \textbf{Initialize:} reaching-destination reward $R_{des}$, out-of-boundary penalty $P_{ob}$, SIR penalty weight, replay memory queue $D$ with capacity $C$ and minibatch size $B$.
			\STATE \textbf{Initialize:} dueling DQN network with coefficients $\pmb{\theta}$, target network with coefficients $\pmb{\theta}^{-}=\pmb{\theta}$.
			\STATE \textbf{Initialize:} SIR map in rain medium obtained from the PWE simulator.
			%			find the maximum obtainable SIR in the whole serving area $\mathcal{S}$, $\text{SIR}$
			\FOR{$n_{epi}=1,\cdots,\overline{N}_{epi}$}
			\STATE \textbf{Initialize:} a sliding window queue $W$ with length $N_1$
			\STATE Randomly set the starting point $\mathbf{q}_{I}\in\mathcal{S}$
			\STATE $\mathbf{q}_0\leftarrow\mathbf{q}_I$, $n\leftarrow0$
			\WHILE{$\lVert\mathbf{q}_n-\mathbf{q}_F\lVert \geq D_{tol}$ \&\& $\mathbf{q}_n\in\mathcal{S}$ \&\& $n\leq \overline{N}_{step}$}
			\STATE Generate a random number $r\sim\mathcal{U}(0,1)$
			\IF{$r<\epsilon$}
			\STATE Action $\overrightarrow{\mathbf{v}}_n\leftarrow\overrightarrow{\mathbf{v}}^{k}, k=\textrm{randi}(K)$
			\ELSE
			\STATE Action $\overrightarrow{\mathbf{v}}_n\leftarrow\overrightarrow{\mathbf{v}}^{k}, k=\underset{k=1,\cdots,K}{\textrm{argmax}}\hat{Q}(\mathbf{q}_n, \overrightarrow{\mathbf{v}}^{(k)};\pmb{\theta})$
			\ENDIF
			\STATE $\mathbf{q}_{n+1}\leftarrow\mathbf{q}+\overrightarrow{\mathbf{v}}_n$; sample maximum SIR  at state $\mathbf{q}_{n+1}$; set the reward $R_n=1-\mu{\text{SIR}}$; store transition $(\mathbf{q}_n, \overrightarrow{\mathbf{v}}_n,R_n,\mathbf{q}_{n+1})$ in the sliding window queue $W$
			%					\algstore{P1}
			%				\end{algorithmic}
			%			\end{algorithm}
			%			
			%			\begin{algorithm}
			%				\caption{Dueling DDQN Multi-Step Learning for Trajectory Design in Rain Medium PII}
			%				\label{alg: DDQNMSRL2}
			%				\begin{algorithmic}[1]
			%					\algrestore{P1}
			\IF{$n\geq N_1$}
			\STATE Calculate the $N_1$-step accumulated return $R_{(n-N_1):n}$ using \eqref{eq: return}. $R_{n+i+1}, i\in{0,\cdots,N_1-1}$ are stored in sliding window, $W$
			\STATE Store transition $(\mathbf{q}_{n-N_1},\overrightarrow{v}_{n-N_1},R_{(n-N_1):n},\mathbf{q}_n)$ in the replay memory $D$
			\ENDIF
			\IF{$|D|\geq B$}
			\STATE Randomly sample a minibatch of $N_1$-step transition $(\mathbf{q}_j,\overrightarrow{\mathbf{v}}_j,R_{j:(j+N_1),\mathbf{q}_{j+N_1}})$ from D
			\FOR{transition $i$ in minibatch}
			\IF{$\lVert \mathbf{q}_{j+N_1}-\mathbf{q}_F \rVert \leq D_{tol}$}
			\STATE $y_j=R_{j:(j+N_1)}+R_{des}$
			\ELSIF{$\mathbf{q}_{j+N_1}\notin\mathcal{S}$}
			\STATE $y_j=R_{j:(j+N_1)}-P_{ob}$
			\ELSE
			\STATE $\overrightarrow{\mathbf{v}}^{*}=\underset{\overrightarrow{\mathbf{v}}^{'}\in\mathcal{A}}{\textrm{agrmax}}\hat{Q}(\mathbf{q}_{j+N_1},\overrightarrow{\mathbf{v}}^{'};\pmb{\theta})$
			\STATE $y_j=R_{j:(j+N_1)} + \gamma^{N_1}\hat{Q}(\mathbf{q}_{j+N_1},\overrightarrow{\mathbf{v}}^{*};\pmb{\theta})$
			\ENDIF
			\ENDFOR
			\STATE Perform a gradient descent step on $\frac{1}{B}\sum^{B-1}_{j=0}(y_j-\hat{Q}(\mathbf{q}_j,\overrightarrow{v}_j;\pmb{\theta}))^2$ w.r.t $\pmb{\theta}$
			\IF{$\textrm{mod}(n,N_{update})==0$}
			\STATE $\pmb{\theta}^{-}\leftarrow\pmb{\theta}$
			\ENDIF
			\ENDIF
			\STATE $n\leftarrow n+1$
			\ENDWHILE
			\ENDFOR
		\end{algorithmic}
	\end{algorithm}
	
\begin{figure*}[!htb]\centering
		%\vspace*{-0.1cm}
		\begin{tikzpicture}
			\node[] at(0,0){\includegraphics[scale=0.44,clip,trim={0cm 0cm 0cm 0cm}]{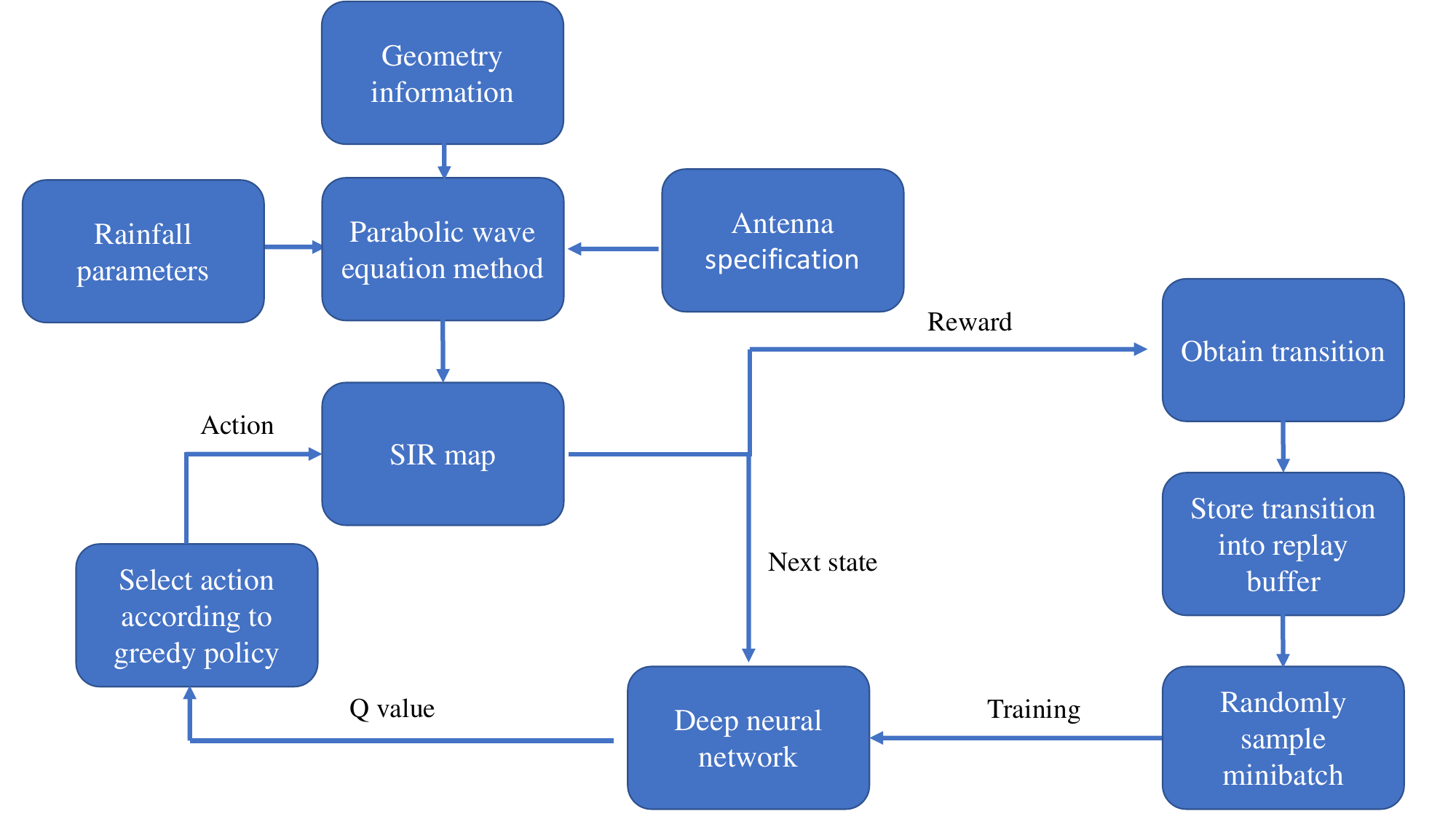}};
		\end{tikzpicture}
		\vspace*{-0.3cm}
		\caption{Framework of the proposed approach of trajectory design for cellular-connected UAVs in rainy environments.}
		\label{fig: Framework}
	\end{figure*}

	The proposed approach for trajectory design in rain medium is summarized in Algorithm 1, and its framework is illustrated in \figref{fig: Framework}.
	The SIR map is calculated by the PWE simulator in Section \ref{Section: rain}, and dueling DDQN multi-step learning is used.
	A sliding window queue with capacity $N_1$ is used to store the $N_1$ latest transitions.
	In particular, setting the maximum number of episodes $\overline{N}_{epi}$ is an important step. This value determines the maximum number of iterations that the agent will interact with the environment and collect data to train the neural network. Once this maximum number of episodes is reached, the training process will stop, and the final DNN parameters will be saved as the trained model. Typically, the maximum number of episodes is chosen based on the complexity of the problem and the available computational resources. A larger number of episodes will generally lead to better performance, but it will also require more time and computational resources. On the other hand, a smaller number of episodes may not be sufficient to learn the optimal policy.
	Therefore, we also investigate the numerical results with various episodes in Section \ref{Section: results}.
	
	\subsection{\re{Computational Complexity Analysis}}
	
	\re{The problem addressed by dueling DDQN is defined by a continuous state space and a discrete action space with $4$ possible actions. During the training process, the agent interacts with the environment, receiving rewards based on its actions and transition probabilities. The computational complexity of the approach primarily stems from two key components: the DNN utilized to approximate the action-value function and the multi-step learning process.}
		
	\re{The computational complexity for computing the action-value function for a single state-action pair through DNN is approximate $O(T * (d + k))$, where $d$ represents the depth (number of layers) of the network, $k$ is the number of steps used for the multi-step update, and $T$ is the number of time steps per episode. Additionally, the multi-step learning technique introduces a computational complexity of $O(b*d)$, where $b$ is the size of each mini-batch. Overall, we estimate the total training computational complexity as $O(N * T * (d + k) + N*b*d)$, where $N$ is the number of episodes.}

	%\vspace{0.3cm}
	
	%%%%%%%%%%%%%%%%%%%%%%%%%%%%%%%%%%%%%%%%%%%%%%%%%%%%%%%%%%%%%%%%%%%%%%%%%%%%%%%%
	%%%%%%%%%%%%%%%%%%%%%%%%%%%%%%%%%%%%%%%%%%%%%%%%%%%%%%%%%%%%%%%%%%%%%%%%%%%%%%%%
	\section{Numerical Results and Performance Evaluation}
	\label{Section: results}
\begin{figure}[!htb]\centering
		\vspace*{-0.1cm}
		\begin{tikzpicture}
			\node[] at(0,0){\includegraphics[scale=0.7,clip,trim={0cm 0cm 0cm 0cm}]{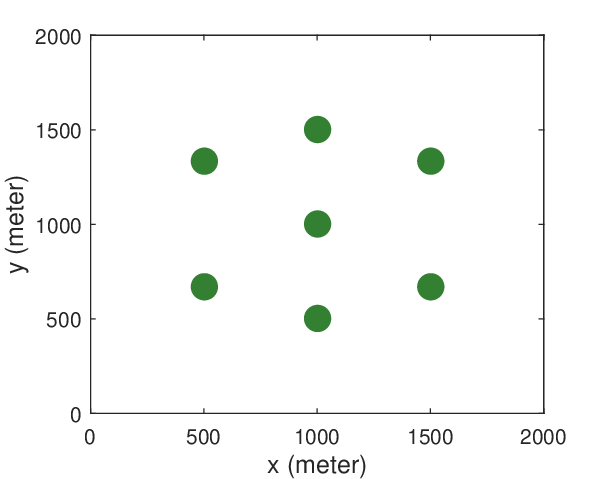}};
		\end{tikzpicture}
		\vspace*{-0.1cm}
		\caption{Locations of the seven ground-based base stations (top view).}
		\label{fig: Simulation_scenario}
	\end{figure}

	To evaluate the effectiveness of our proposed approach, in this section, we present numerical results from a 2\,km$\times$2\,km airspace. 
	During testing, the UAV maintains a fixed altitude of 100\,m while flying. The considered scenario includes seven ground-based base stations, whose locations are illustrated in \figref{fig: Simulation_scenario}.
	We generate the signal-to-interference ratio (SIR) map using the PWE simulator, taking into account the effects of rainfall. 
	Specifically, we set the rain rate to $R = 25$\,mm/h. For simplicity, we utilize a unit-strength Gaussian beam as the radiating source at each base station, and the operating frequency is set to 4.9\,GHz.	
	%	The expression of initial field distributions is given as 
	%	\begin{equation}
	%		\begin{aligned}
	%			\label{eq:10}
	%			&u_0\left(x,y\right)=\\
	%			&\text{exp}\left\{-jk_0x\sin{\theta_{eax}}\ -\frac{\left[k_0\sin{\left(0.5\theta_{bwx}\right)}\left(x\ -\ x_s\right)\right]^2}{2\ln{2}}\right\}\\
	%			&\text{exp}\left\{-jk_0y\sin{\theta_{eay}}\ -\frac{\left[k_0\sin{\left(0.5\theta_{bwy}\right)}\left(y\ -\ y_s\right)\right]^2}{2\ln{2}}\right\}
	%		\end{aligned}
	%	\end{equation}
	%	Where $\theta_{eax}$, $\theta_{eay}$, $\theta_{bwx}$, $\theta_{bwy}$ denote the elevation angle and 3dB beamwidth in $x$ and $y$ direction. In this section, the parameter is set as $\theta_{eax}=\theta_{eay}=5^{\circ}$, $\theta_{bwx}=\ \theta_{bwy}=4.333^{\circ}$.
	
\begin{figure}[!htb]\centering
		\vspace*{-0.5cm}
		\hspace*{-1mm}
		\subfloat[][]{
			\begin{tikzpicture}
				\node[] at(0,0){\includegraphics[scale=0.56,clip,trim={0cm 0cm 0cm 0cm}]{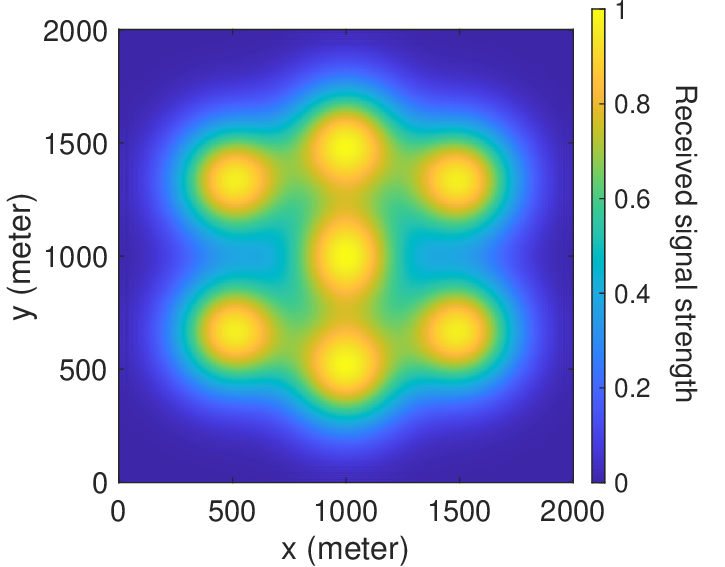}};
			\end{tikzpicture}
		}
		\vspace*{-0.2cm}
		
		\subfloat[][]{
			\begin{tikzpicture}
				\node[] at(0,0){\includegraphics[scale=0.56,clip,trim={0cm 0cm 0cm 0cm}]{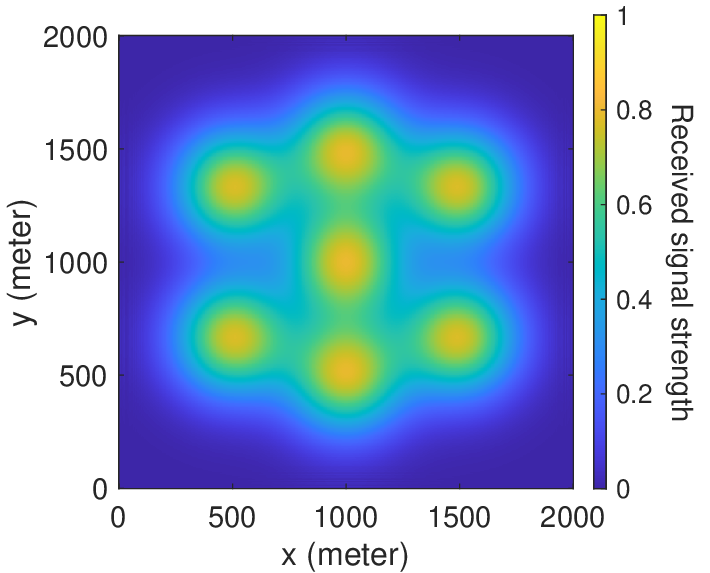}};
			\end{tikzpicture}
		}
		\vspace*{-0.1cm}
		\caption{Received signal strength in the airspace at a fixed height of 100\,m. (a) Homogeneous atmosphere medium. (b) Rain medium.}
		\label{fig: RSS_J}
	\end{figure}

	First, we use the PWE simulator to obtain the associated RSS in such an environment with varying weather conditions. Specifically, we examine a homogeneous atmosphere medium and a rain medium. 
	The RSS distribution on a horizontal plane at a height of 100\,m in the airspace is compared in \figref{fig: RSS_J}. 
	%Subsequently, SIR in the airspace can be calculated using \eqref{eq: SIR} and the corresponding SIR map is illustrated in \figref{fig: SIR_Map_v4}.
	The results show that the rainy environment has a significant impact on radio wave propagation. 
	Therefore, accurate physics-based wave propagation models are necessary for UAV trajectory design under such weather conditions.	
	
	%\begin{figure}[!htb]\centering
	%%\vspace*{-0.25cm}
	%\hspace*{0.4 cm}
	%\subfloat[][]{
	%\begin{tikzpicture}
	%\node[] at(0,0){\includegraphics[scale=0.44,clip,trim={0cm 0.4cm 0cm 0.2cm}]{CoverageMapTrue_Norain.pdf}};
	%\end{tikzpicture}
	%}
	%%		\vspace*{-5.1mm}
	%\hspace*{-0.7 cm}
	%\subfloat[][]{
	%\begin{tikzpicture}
	%\node[] at(0,0){\includegraphics[scale=0.44,clip,trim={0cm 0.4cm 0cm 0.2cm}]{CoverageMapTrueRain1.pdf}};
	%\end{tikzpicture}
	%}
	%\caption{Signal-to-interference ratio in the airspace at a fixed height of 100\,m. (a) Homogeneous atmosphere medium. (b) Rain medium.}
	%\label{fig: SIR_Map_v4}
	%\end{figure}
	
\begin{figure}[!htb]\centering
		\vspace*{-0.5cm}

		\subfloat[][]{
			\begin{tikzpicture}
				\node[] at(0,0){\includegraphics[scale=0.47,clip,trim={0cm 0cm 0cm 0cm}]{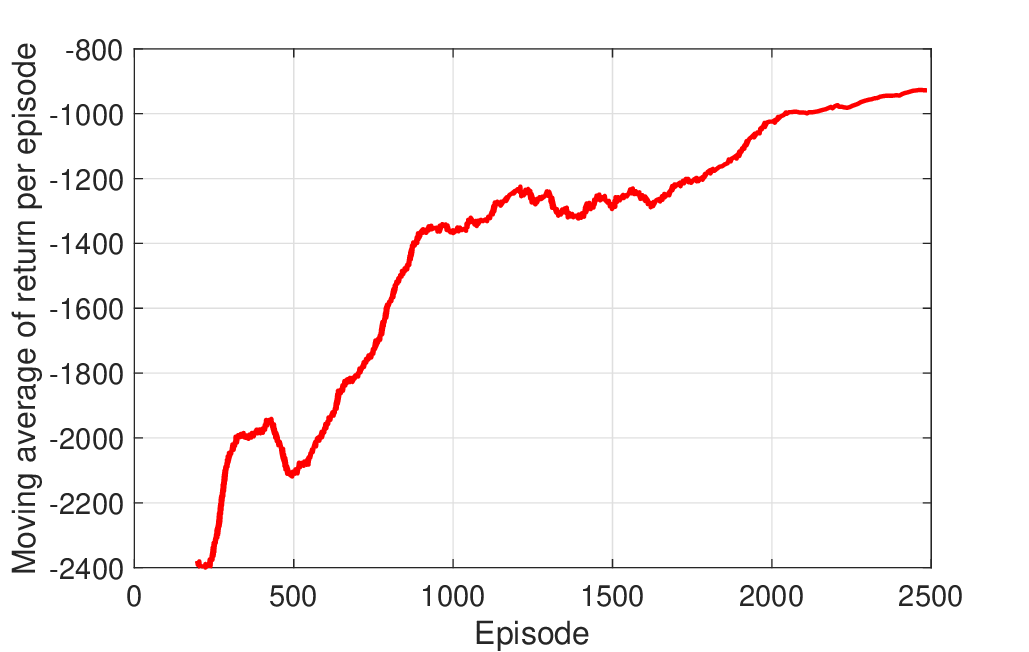}};
			\end{tikzpicture}
		}
		%\vspace*{-5.1mm}
		\vspace*{-0.5cm}
		
		\subfloat[][]{
			\begin{tikzpicture}
				\node[] at(0,0){\includegraphics[scale=0.47,clip,trim={0cm 0cm 0cm 0cm}]{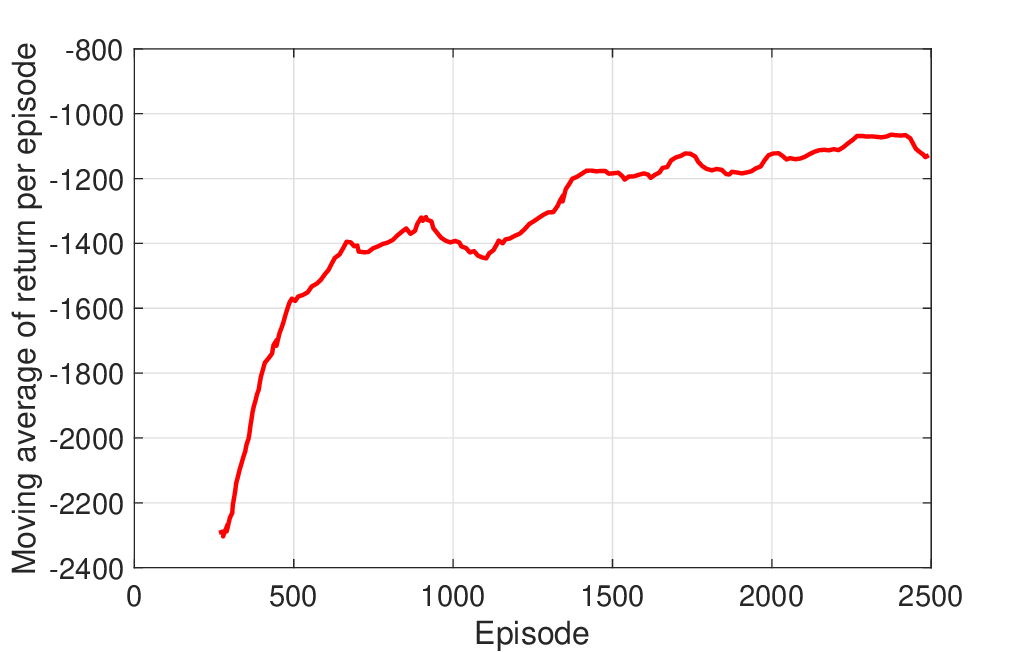}};
			\end{tikzpicture}
		}
		%\vspace*{-0.1cm}
		\caption{Moving average return. (a) Homogeneous atmosphere medium. (b) Rain medium.}
		\label{fig: Return_conf}
	\end{figure}

\begin{figure*}[!htb]\centering
		%\vspace*{-1.1cm}
		\subfloat[][]{
			\begin{tikzpicture}
				\node[] at(0,0){\includegraphics[scale=0.38,clip,trim={0cm 0.35cm 0cm 0.2cm}]{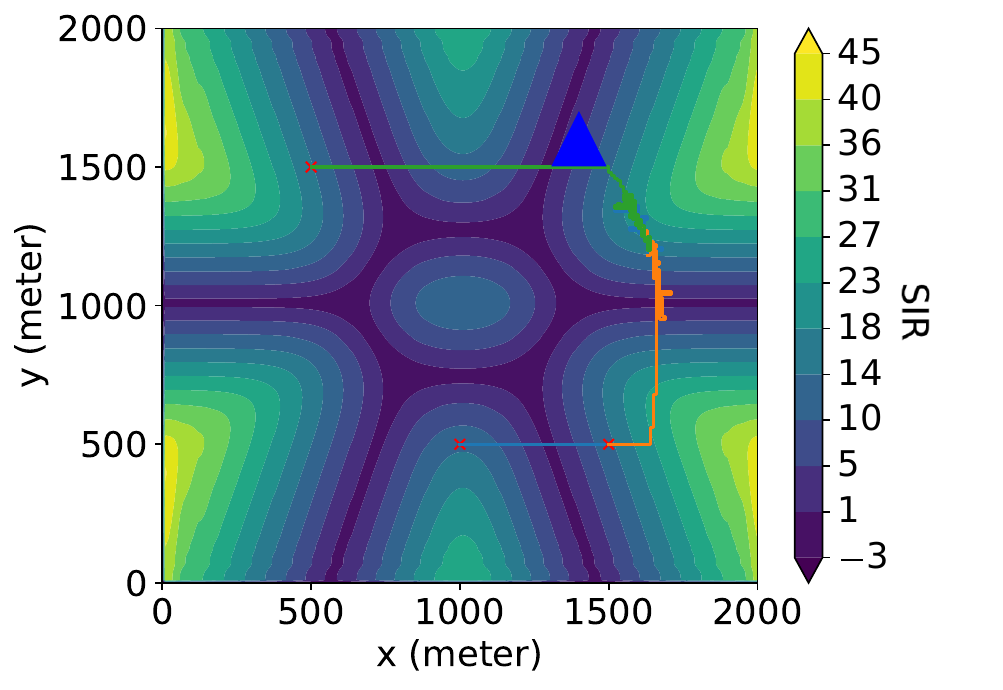}};
			\end{tikzpicture}
		}
		\hspace*{-0.9 cm}
		\subfloat[][]{
			\begin{tikzpicture}
				\node[] at(0,0){\includegraphics[scale=0.38,clip,trim={0cm 0.35cm 0cm 0.2cm}]{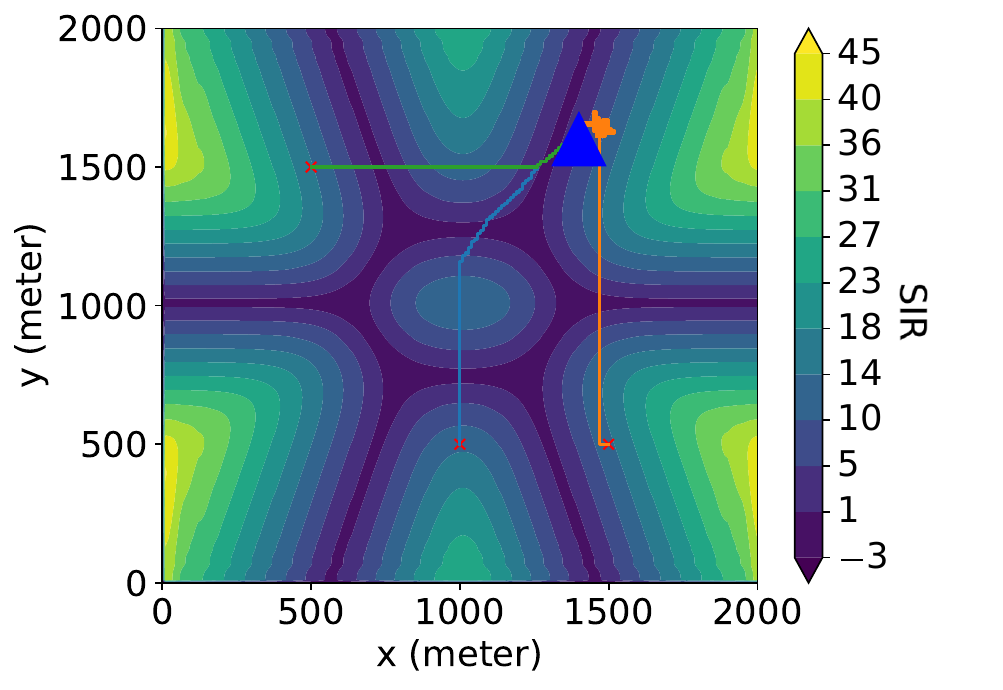}};
			\end{tikzpicture}
		}
		\hspace*{-0.9 cm}
		\subfloat[][]{
			\begin{tikzpicture}
				\node[] at(0,0){\includegraphics[scale=0.38,clip,trim={0cm 0.35cm 0cm 0.2cm}]{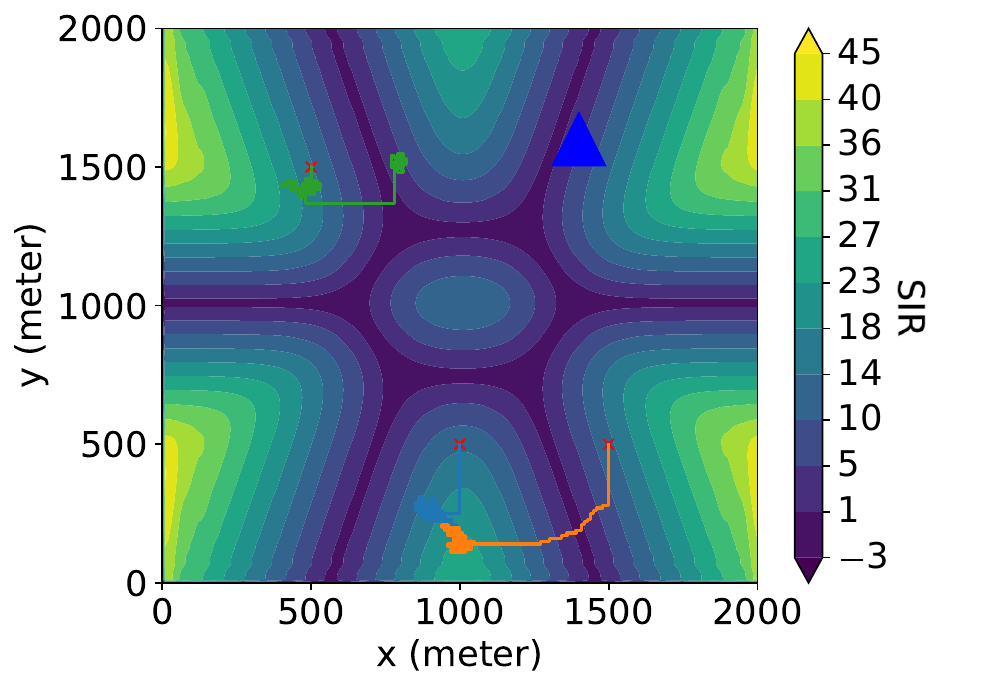}};
			\end{tikzpicture}
		}

		\vspace*{-5.2mm}
		\subfloat[][]{
			\begin{tikzpicture}
				\node[] at(0,0){\includegraphics[scale=0.38,clip,trim={0cm 0.35cm 0cm 0.2cm}]{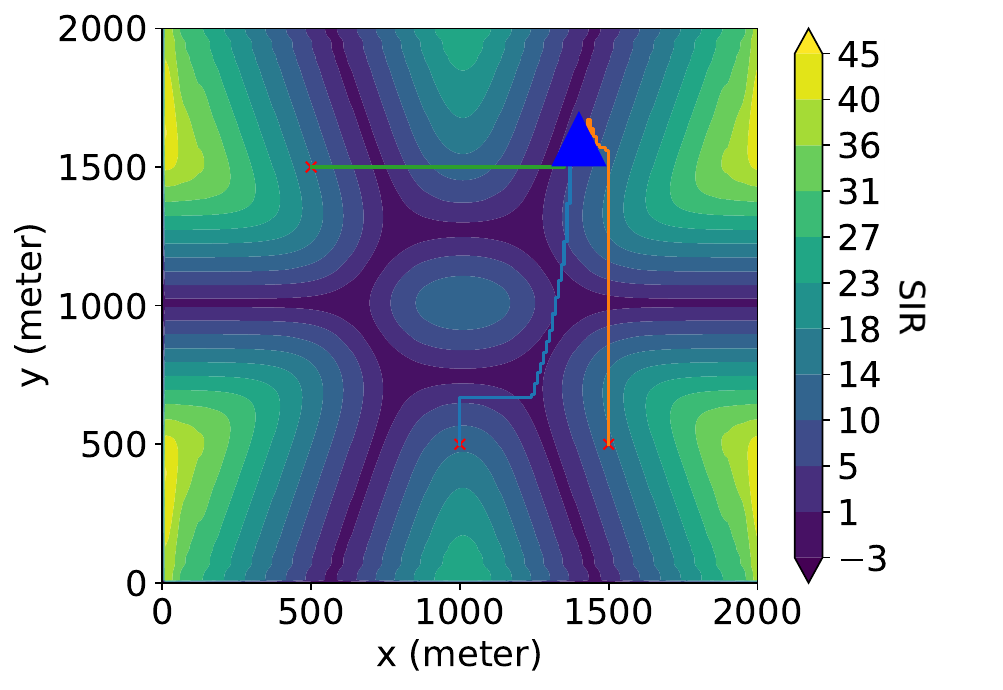}};
			\end{tikzpicture}
		}
		\hspace*{-0.9 cm}
		\subfloat[][]{
			\begin{tikzpicture}
				\node[] at(0,0){\includegraphics[scale=0.38,clip,trim={0cm 0.35cm 0cm 0.2cm}]{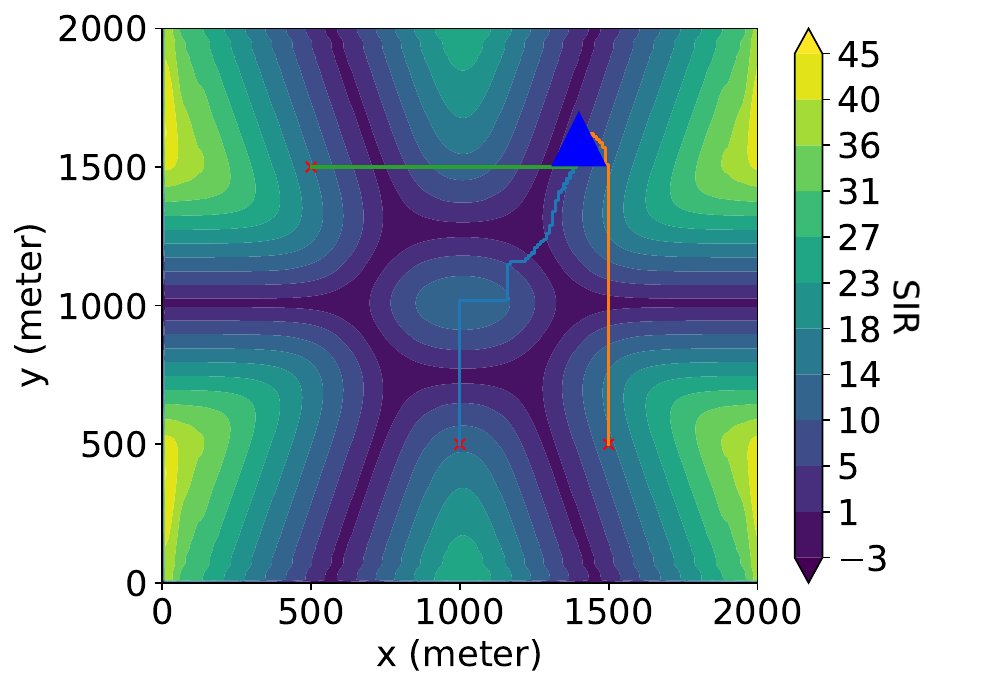}};
			\end{tikzpicture}
		}
		\hspace*{-0.9 cm}
		\subfloat[][]{
			\begin{tikzpicture}
				\node[] at(0,0){\includegraphics[scale=0.38,clip,trim={0cm 0.35cm 0cm 0.2cm}]{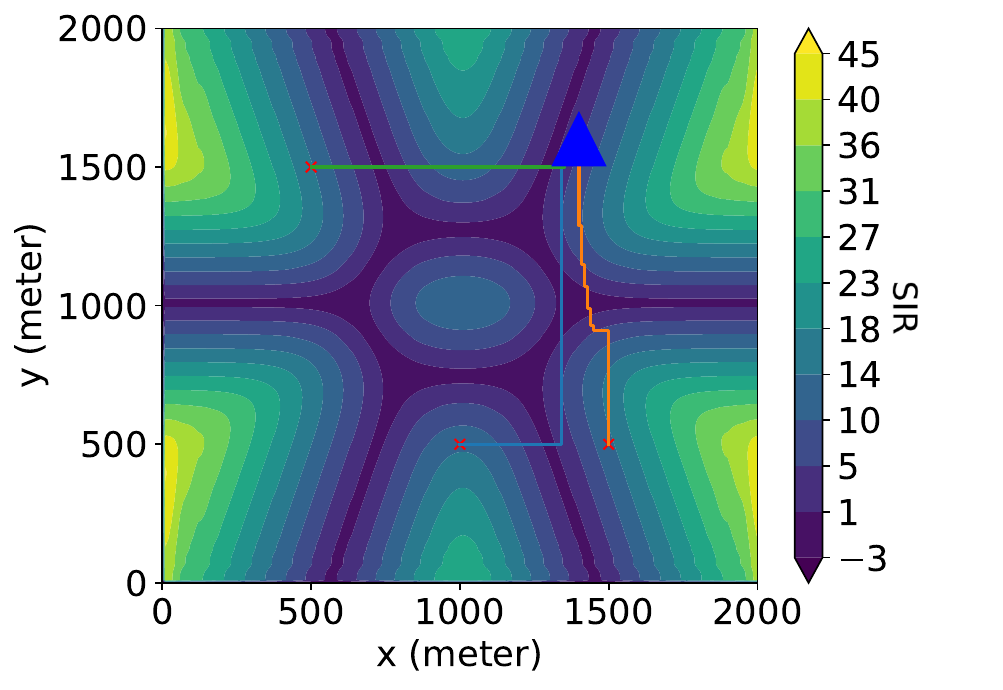}};
			\end{tikzpicture}
		}
	
		%	\vspace*{-4.1mm}
		\vspace*{-5.2mm}
		\subfloat[][]{
			\begin{tikzpicture}
				\node[] at(0,0){\includegraphics[scale=0.38,clip,trim={0cm 0.35cm 0cm 0.2cm}]{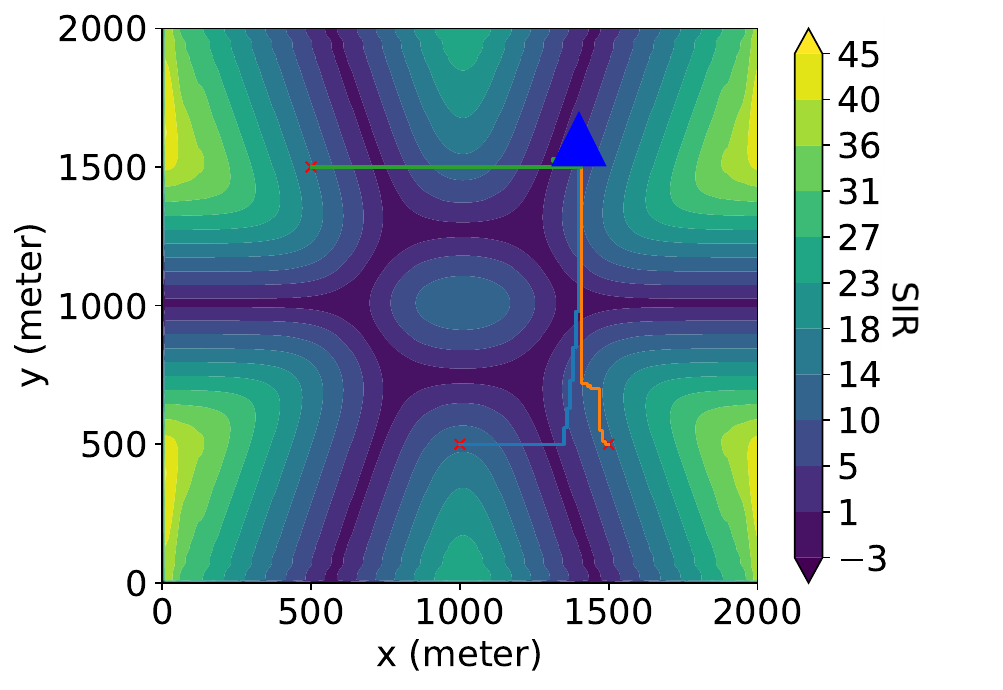}};
			\end{tikzpicture}
		}
		\hspace*{-0.9 cm}
		\subfloat[][]{
			\begin{tikzpicture}
				\node[] at(0,0){\includegraphics[scale=0.38,clip,trim={0cm 0.35cm 0cm 0.2cm}]{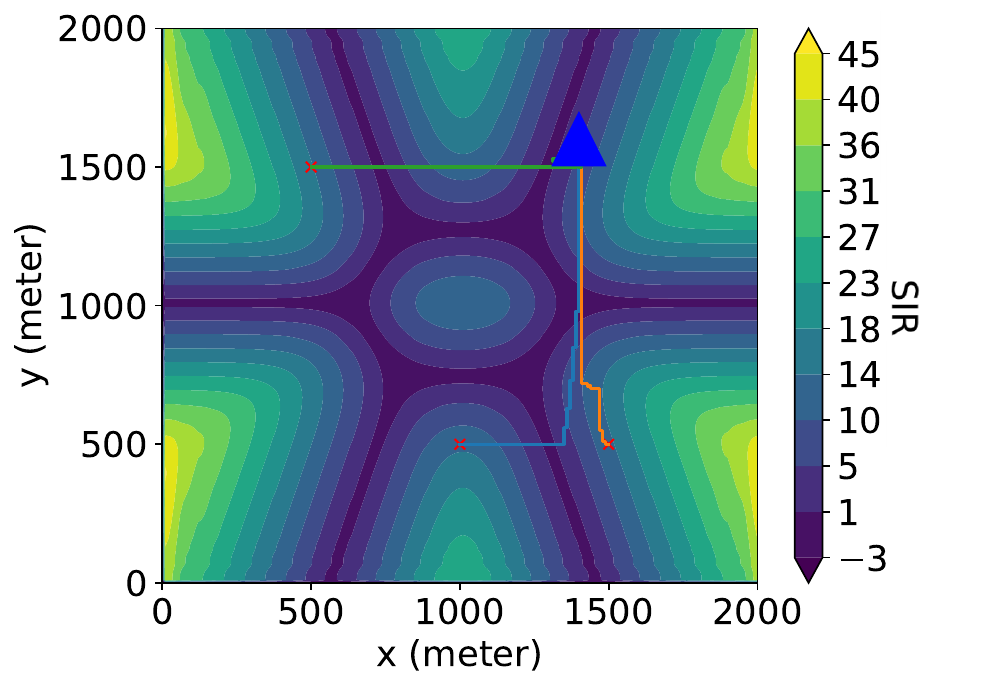}};
			\end{tikzpicture}
		}
		\hspace*{-0.9 cm}
		\subfloat[][]{
			\begin{tikzpicture}
				\node[] at(0,0){\includegraphics[scale=0.38,clip,trim={0cm 0.35cm 0cm 0.2cm}]{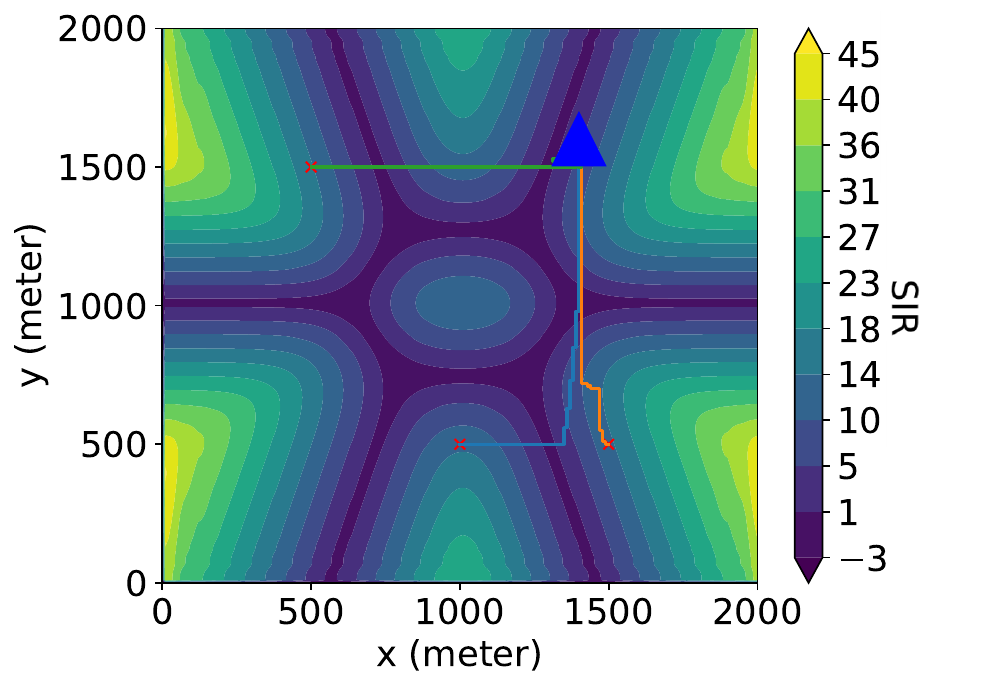}};
			\end{tikzpicture}
		}
		
		\vspace*{-0.05cm}
		%\vspace*{0.5cm}
		\caption{Trajectories at different episodes during the training process. (a) 600 episodes. (b) 900 episodes. (c) 1200 episodes. (d) 1500 episodes. (e) 1800 episodes. (f) 2100 episodes. (g) 2400 episodes. (h) 2700 episodes. (i) 3000 episodes.}
		\label{fig: Trajectory_during_train}
	\end{figure*}

	Second, dueling DDQN with multi-step learning is employed to optimize the UAV's trajectory.
	Specifically, we randomly select 200 initial starting points and set a fixed destination point.
	During the training process, the moving average return of the proposed algorithm is presented in \figref{fig: Return_conf}.
	Additionally, the trajectories at different episodes of three initial points with rain are presented in \figref{fig: Trajectory_during_train}. 
	The initial points are marked using red crosses, while the blue triangle indicates the final destination point. 
	\re{The minimum SIR is set as 10\,dB in this section.}
	It can be seen that the proposed algorithm converges as the number of episodes increases.
	
\begin{figure}[!htb]\centering
		%\vspace*{-0.3cm}
%		\hspace*{0.4 cm}
		\subfloat[][]{
			\begin{tikzpicture}
				\node[] at(0,0){\includegraphics[scale=0.45,clip,trim={0cm 0.38cm 0cm 0cm}]{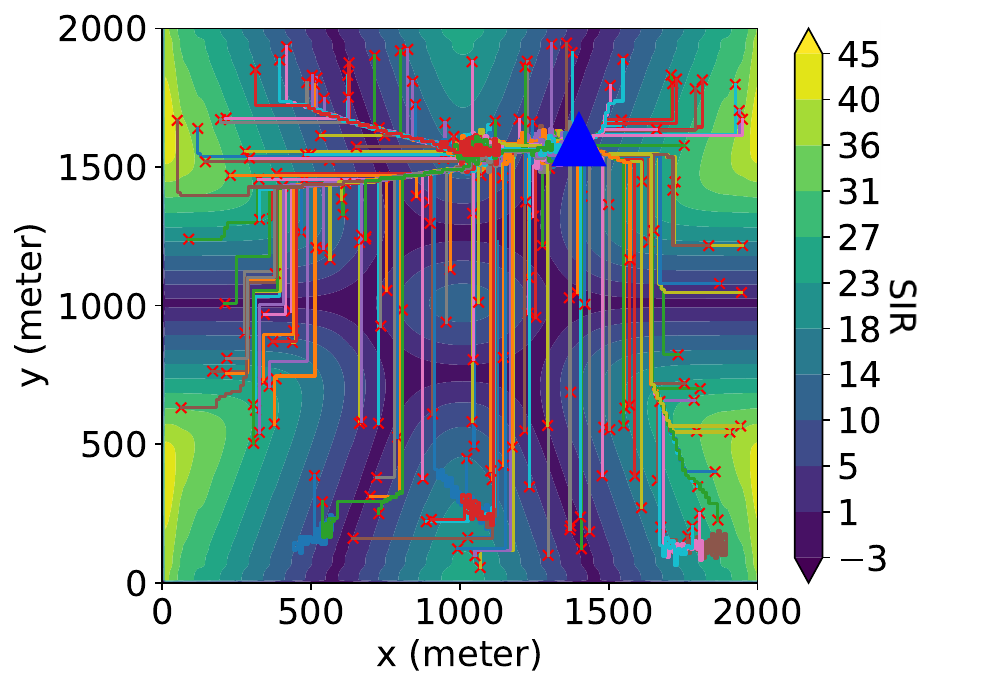}};
			\end{tikzpicture}
		}
		\vspace*{-0.5cm}
		\hspace*{0cm}
		
		\subfloat[][]{
			\begin{tikzpicture}
				\node[] at(0,0){\includegraphics[scale=0.45,clip,trim={0cm 0.38cm 0cm 0cm}]{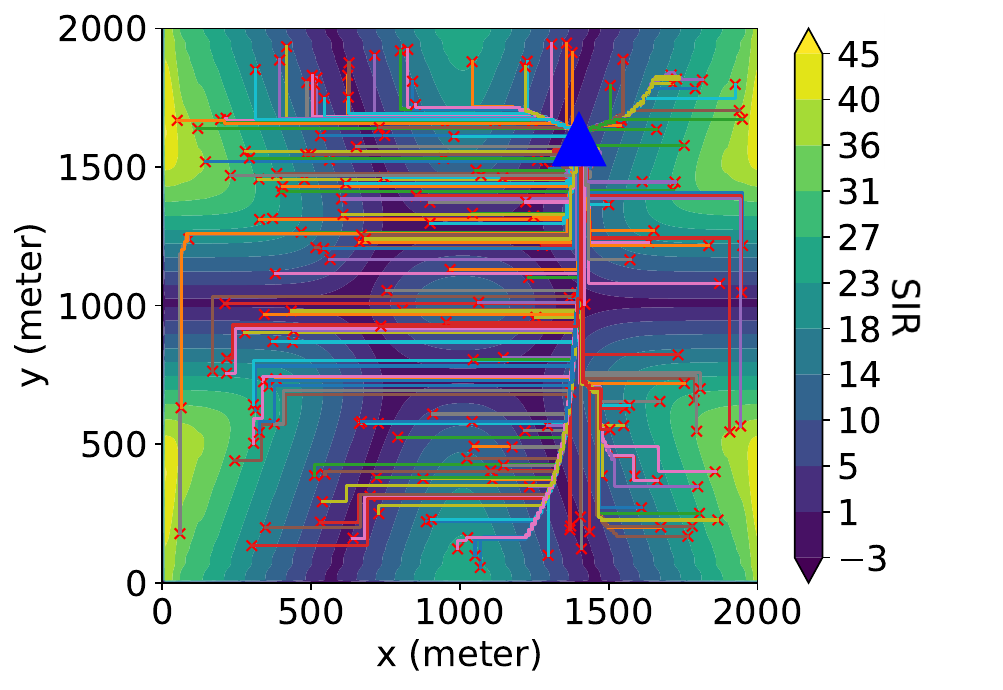}};
			\end{tikzpicture}
		}
		
		\vspace*{-4.8mm}
		\vspace*{0.3cm}
		\caption{Designed trajectories in the airspace at a fixed height of 100\,m. (a) Homogeneous atmosphere medium. (b) Rain medium.}
		\label{fig: Trajectory_without_rain}
	\end{figure}

	Upon convergence, the designed trajectories for the cellular-connected UAVs in the homogeneous atmosphere medium and the rain medium are plotted in \figref{fig: Trajectory_without_rain}. 
	The parameters of the DRL algorithm in this section are summarized in Table~\ref{table: simuParas}.
	Note that the same set of initial locations is used for the two cases with different mediums in \figref{fig: Trajectory_without_rain}. 
	\re{The proposed model is trained on an NVIDIA Geforce RTX 3090 GPU, and the execution time is 3480\,s.}
	The results show that the UAVs can avoid weak coverage regions and reach the destination point successfully.
	Additionally, the optimal trajectories for the same initial point differ between the two cases, indicating that the presence of rain has affected the trajectory.
	%The parameters of the DRL algorithm in this section are summarized in Table~\ref{table: simuParas}.
	%% Table %%%%%%%%%%%%%%%%%%%%%%%%%%%%%%%%%%%%%%%%%%%%%%%%%%%%%%%%%%%%%%%%%%%%%%%%%%%%
	\begin{table}[!htb]
		\begin{center}
			\caption{Parameters for the Proposed Trajectory Design Algorithm}
			%\vspace*{1em}
			\label{table: simuParas} 
			\renewcommand\arraystretch{2.5}
			%\begin{tabular}{|c|p{3cm}<\centering|c|c|}\hline
			\begin{tabular}{|m{3.4cm}<\centering|m{1.9cm}<\centering|m{1.9cm}<\centering|}\hline
				\textbf{Parameter Meaning} & \textbf{Symbol} & \textbf{Value} \\
				\hline
				maximum number of episodes &  $\overline{N}_{epi}$ & $3000$\\
				\hline
				maximum number of steps per episode & $\overline{N}_{step}$ & $300$\\
				\hline
				update frequency & $N_{update}$ & $5$\\
				\hline
				reaching-destination toleration distance  & $D_{tol}$ & $10$\\
				\hline
				initial exploration & $\epsilon_{0}$ & $0.5$\\
				\hline
				decaying rate & $\alpha$ & $0.998$\\
				\hline
				reaching-destination reward & $R_{des}$ & $2000$\\
				\hline
				out-of-boundary penalty & $P_{op}$ & $10000$\\
				\hline
				SIR penalty weight & $\mu$ & $\frac{10}{43}$\\
				\hline
				replay memory queue capacity & $C$ & $100000$\\
				\hline
				minibatch size & $B$ & $16$\\
				\hline
			\end{tabular}
		\end{center}
		%		\vspace*{5pt}
		%\hrulefill
		%\vspace*{-5pt}
	\end{table}
	%% Table %%%%%%%%%%%%%%%%%%%%%%%%%%%%%%%%%%%%%%%%%%%%%%%%%%%%%%%%%%%%%%%%%%%%%%%%%%%%
	
\begin{figure}[!htb]\centering
		%\vspace*{-0.3cm}
%		\hspace*{0.4 cm}
		\subfloat[][]{
			\begin{tikzpicture}
				\node[] at(0,0){\includegraphics[scale=0.45,clip,trim={0cm 0.38cm 0cm 0cm}]{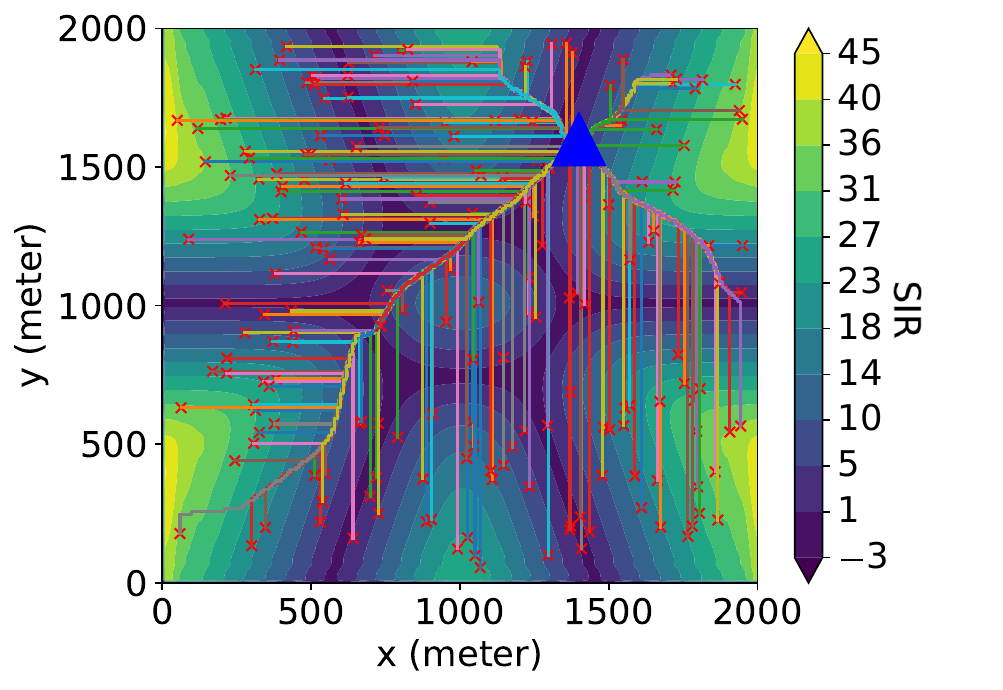}};
			\end{tikzpicture}
		}
		\vspace*{-0.5cm}
		\hspace*{0cm}
		
		\subfloat[][]{
			\begin{tikzpicture}
				\node[] at(0,0){\includegraphics[scale=0.45,clip,trim={0cm 0.38cm 0cm 0cm}]{trajectoriesNoMapping_Rain1trained200.pdf}};
			\end{tikzpicture}
		}
		
		\vspace*{-4.8mm}
		\vspace*{0.3cm}
		\caption{Designed trajectories in the airspace at a fixed height of 100\,m in rain medium with different $\mu$. (a) $\mu = 0.1$. (b) $\mu = 10$.}
		\label{fig: Trajectory_with_rain}
	\end{figure}

	Subsequently, we investigate the impact of two key parameters, namely, the weight factor, $\mu$, and the rain rate, $R$, on the UAV trajectory design. 
	In \figref{fig: Trajectory_with_rain}, UAV trajectories in cases with two different values of $\mu$ are plotted. \re{The execution time of two cases with different $\mu$ is 3254\,s and 3252\,s, respectively.}
	As can be seen, when $\mu = 0.1$, the UAVs focus more on the flying time, and prefer to reach the destination point quickly. In contrast, when $\mu = 10$, the UAVs can better avoid the weak coverage regions. Thus, the choice of $\mu$ can be based on specific mission requirements. 
	
\begin{figure}[!htb]\centering
%		\vspace*{-0.1cm}
		%		\hspace*{-2mm}
		%\hspace*{0.5 cm}
		\subfloat[][]{
			\begin{tikzpicture}
				\node[] at(0,0){\includegraphics[scale=0.45,clip,trim={0cm 0.3cm 0cm 0cm}]{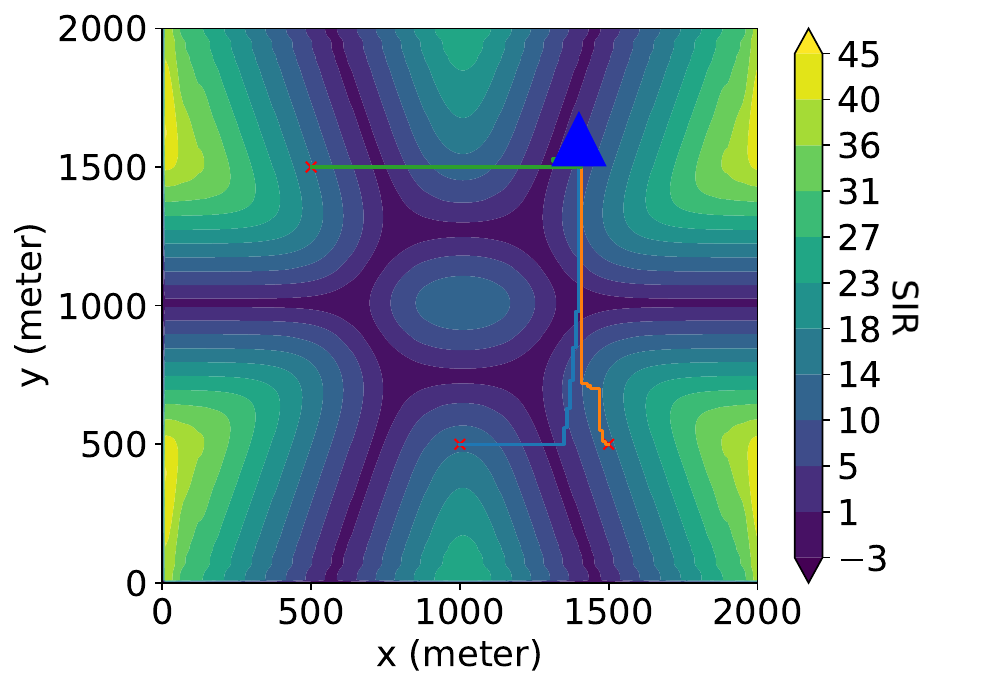}};
			\end{tikzpicture}
		}

		%\hspace*{0.5 cm}
		\vspace*{-5.1mm}
		\subfloat[][]{
			\begin{tikzpicture}
				\node[] at(0,0){\includegraphics[scale=0.45,clip,trim={0cm 0.3cm 0cm 0cm}]{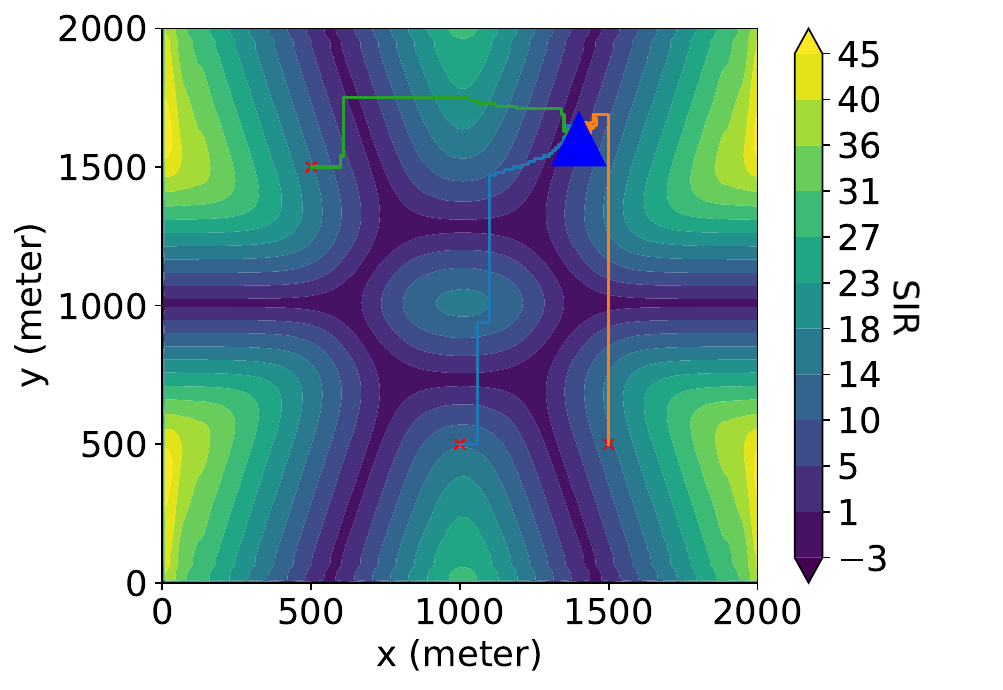}};
			\end{tikzpicture}
		}
%		\hspace*{5mm}
		
		\vspace*{-5.1mm}
		\subfloat[][]{
			\begin{tikzpicture}
				\node[] at(0,0){\includegraphics[scale=0.45,clip,trim={0cm 0.3cm 0cm 0cm}]{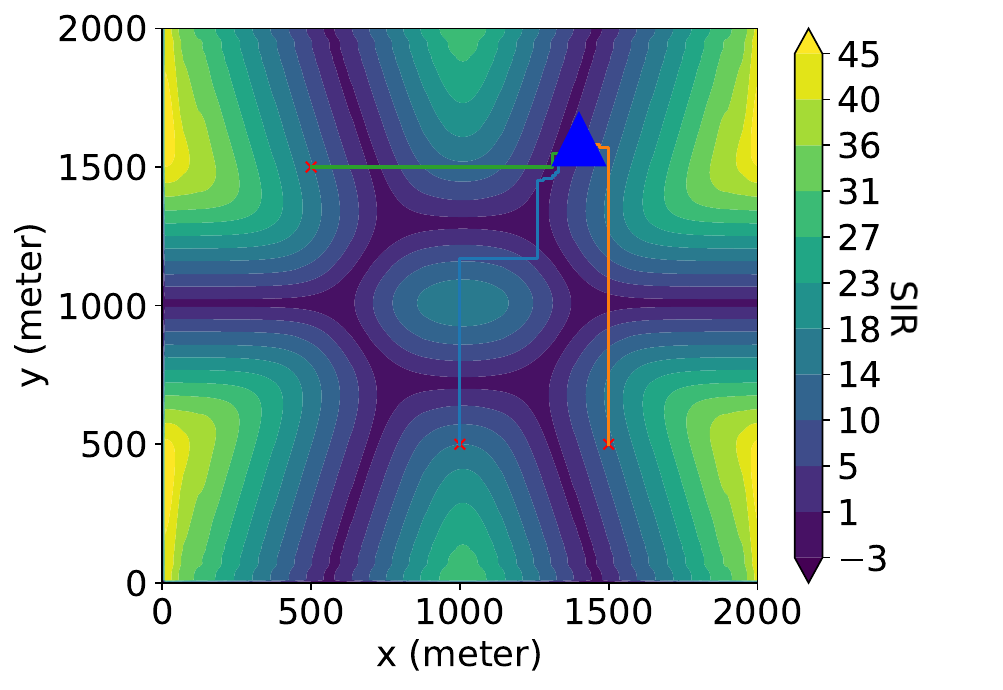}};
			\end{tikzpicture}
		}
		%	\vspace*{-6.1mm}
		%\vspace*{1mm}
		\caption{Designed trajectories in the airspace at a fixed height of 100\,m with different rain rates. (a) $R = 25$\,mm/h. (b) $R = 50$\,mm/h. (c) $R = 100$\,mm/h.}
		\label{fig: Trajectory_with_different_rain}
	\end{figure}

	To evaluate the impact of various rainy weather conditions on the UAV trajectory design, we conduct experiments with different rain rates. 
	%	In order to comprehensively evaluate the performance of our proposed algorithm, we conducted experiments under various rain conditions.
	Specifically, UAV trajectories under different rain rates $R$ are we generated and compared in \figref{fig: Trajectory_with_different_rain}.
	We consider three distinct cases with rain rates set to $R = 25$\,mm/h, $R = 50$\,mm/h, and $R = 100$\,mm/h, respectively. 
	\re{The execution time of three cases with different rain rates is 3194\,s, 3208\,s, and 3201\,s, respectively.}
	The experimental results demonstrate that our approach is capable of effectively handling different levels of rain, indicating its robustness and adaptability in practical scenarios with diverse weather conditions.
	
\begin{figure}[!htb]\centering
%		\vspace*{-0.1cm}
		%		\hspace*{-2mm}
		%\hspace*{0.5 cm}
		\subfloat[][]{
			\begin{tikzpicture}
				\node[] at(0,0){\includegraphics[scale=0.45,clip,trim={0cm 0.38cm 0cm 0cm}]{Rain1_3points.pdf}};
			\end{tikzpicture}
		}
		\vspace*{-5.1mm}
		
		\hspace*{-0.5cm}
		\subfloat[][]{
			\begin{tikzpicture}
				\node[] at(0,0){\includegraphics[scale=0.45,clip,trim={0cm 0.38cm 0cm 0cm}]{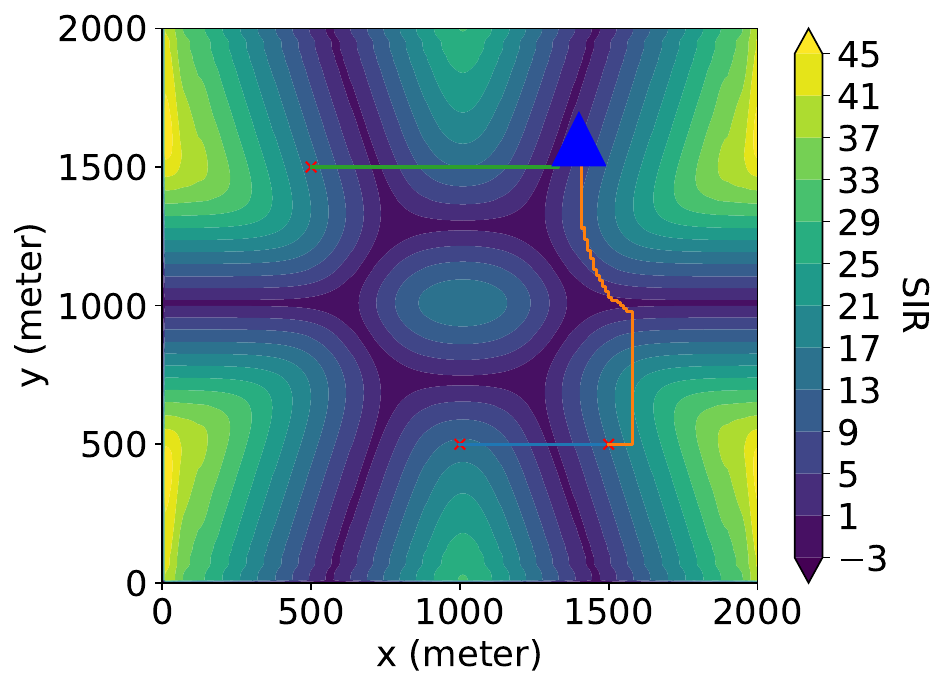}};
			\end{tikzpicture}
		}
	
		\vspace*{-5.1mm}
	\subfloat[][]{
		\begin{tikzpicture}
			\node[] at(0,0){\includegraphics[scale=0.45,clip,trim={0cm 0.3cm 0cm 0cm}]{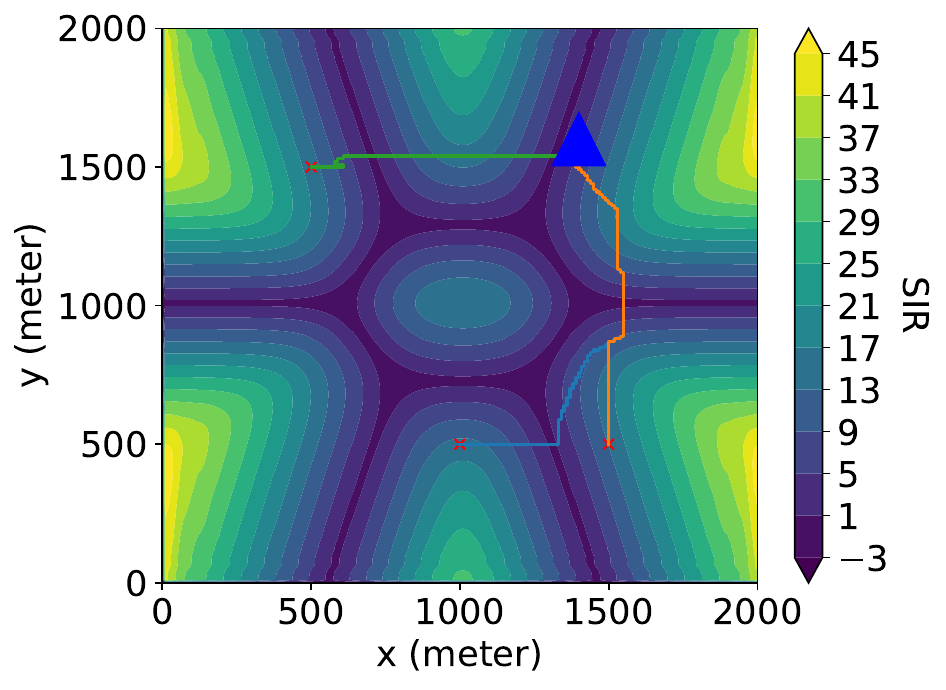}};
		\end{tikzpicture}
	}
		%	\vspace*{-6.1mm}
		%\vspace*{1mm}
		\caption{\re{Designed trajectories in the airspace at a fixed height of 100\,m. (a) Rain medium without wind condition. (b) Rain medium with wind condition (wind speed: 2\,m/s). (c) Rain medium with wind condition (wind speed: 5\,m/s).}}
		\label{fig: Trajectory_with_different_wind}
	\end{figure}

	\re{Additionally, the wind in a rainy medium can also affect wave propagation mainly due to the wind-induced movement of raindrops in a rainy medium introduces variations in the scattering environment~\cite{Asen03}. A correction factor is employed by considering the individual drop-size dependent velocity resultant:}
	\begin{equation}
		\re{F(D) = \cos{\left(\tan^{-1}{\left(\frac{v_h}{v_v(D)}\right)}\right)}}
	\end{equation}
	\re{where $v_h$ is the horizontal wind speed. $v_v(D)$ indicates the drop diameter dependent terminal velocity and can be found in~\cite{Brussaard94}.}
	\re{Then, the effective permittivity in rain medium \eqref{eq: eff_rain} can be obtained while considering the wind condition:}
	\begin{equation}
		\begin{split}
			&\re{\varepsilon_{\text{eff}} = \varepsilon_{0} + \int_{D_\text{min}}^{1.25} N(D)4\pi \frac{\varepsilon_{0}(\varepsilon_{\omega}-\varepsilon_{0})}{\varepsilon_{\omega}-2\varepsilon_{0}}(\frac{D}{2})^3}\\
			&\re{\times\frac{3\varepsilon_{0} / F(D_i)}{3\varepsilon_{0}-4\pi \displaystyle{\frac{\varepsilon_{0}(\varepsilon_{\omega}-\varepsilon_{0})}{\varepsilon_{\omega}-2\varepsilon_{0}}}(\frac{D}{2})^3} dD}\\
			&\re{+ \frac{1}{3}\sum_{i=1}^{3}\int_{1.25}^{D_\text{max}} N(D)\frac{4\pi}{3}\frac{\varepsilon_{0}(\varepsilon_{\omega}-\varepsilon_{0})}{\varepsilon_{\omega}+L_i(\varepsilon_{\omega}-\varepsilon_{0})} (\frac{D}{2})^3 }\\
			&\re{\times \frac{\varepsilon_{0}/F(D_i)}{\varepsilon_{0}-L_i \displaystyle{\frac{4\pi}{3}\frac{\varepsilon_{0}(\varepsilon_{\omega}-\varepsilon_{0})}{\varepsilon_{\omega}+L_i(\varepsilon_{\omega}-\varepsilon_{0})}} (\frac{D}{2})^3} dD.}
		\end{split}
		\label{eq: eff_wind}
	\end{equation}
	\re{In \figref{fig: Trajectory_with_different_wind}, optimized trajectories considering about the wind condition are presented. The rain rates are all $R = 25$\,mm/h. In the two cases considering the wind condition, the horizontal wind speed is set as $v_v(D)=2$ m/s and $v_v(D)=5$ m/s, respectively. The proposed model takes approximately 3192\,s to run in the case that accounts for the wind condition. It can be seen that the variations in the wind can affect radio wave propagation, leading to changes in the SIR distribution. Consequently, the UAVs' optimized trajectories need to be adjusted accordingly.}

	%%%%%%%%%%%%%%%%%%%%%%%%%%%%%%%%%%%%%%%%%%%%%%%%%%%%%%%%%%%%%%%%%%%%%%%%%%%%%%%%
	%%%%%%%%%%%%%%%%%%%%%%%%%%%%%%%%%%%%%%%%%%%%%%%%%%%%%%%%%%%%%%%%%%%%%%%%%%%%%%%%
	\section{Conclusion}
	\label{Section: conclusion}
	This paper presents a novel physics-based trajectory design approach for cellular-connected UAVs in rainy environments. Compared to the previous optimal trajectory studies, the proposed approach takes into account detailed information about the environment's geometry and the impact of weather conditions on radio wave propagation and trajectory design. To formulate the trajectory design problem, we have defined the state, action space, and reward function, and a site-specific electromagnetic simulator and dueling DDQN with multi-step learning are utilized to solve this problem. The comparison of the optimal trajectory for cellular-connected UAVs in the atmosphere medium and rain medium demonstrates the necessity of our proposed approach. Additionally, we have provided a thorough study of varying weather conditions on trajectory design and investigated the impact of various parameters in the formulated problem. Overall, the proposed approach provides a valuable contribution to the field of UAV trajectory design in complex and challenging weather environments and has great potential for various UAV applications, such as weather monitoring and search and rescue operations.

	%\vspace*{12cm}
	%%%%%%%%%%%%%%%%%%%%%%%%%%%%%%%%%%%%%%%%%%%%%%%%%%%%%%%%%%%%%%%%%%%%%%%%%%%%%%%%
	%%%%%%%%%%%%%%%%%%%%%%%%%%%%%%%%%%%%%%%%%%%%%%%%%%%%%%%%%%%%%%%%%%%%%%%%%%%%%%%%
	%% Appendix
	% if have a single appendix: \appendix[Proof of the Zonklar Equations]
	% or \appendix  % for no appendix heading
	

	%\appendices
	%% you can choose not to have a title for an appendix
	%% if you want by leaving the argument blank
	%\section{}

	%%%%%%%%%%%%%%%%%%%%%%%%%%%%%%%%%%%%%%%%%%%%%%%%%%%%%%%%%%%%%%%%%%%%%%%%%%%%%%%%
	%%%%%%%%%%%%%%%%%%%%%%%%%%%%%%%%%%%%%%%%%%%%%%%%%%%%%%%%%%%%%%%%%%%%%%%%%%%%%%%%
	%% Acknowledgement
	%% use section* for acknowledgement
	%\section*{Acknowledgment}

	%%%%%%%%%%%%%%%%%%%%%%%%%%%%%%%%%%%%%%%%%%%%%%%%%%%%%%%%%%%%%%%%%%%%%%%%%%%%%%%%
	%%%%%%%%%%%%%%%%%%%%%%%%%%%%%%%%%%%%%%%%%%%%%%%%%%%%%%%%%%%%%%%%%%%%%%%%%%%%%%%%
	%% References
	\ifCLASSOPTIONcaptionsoff
	\newpage
	\fi
	
	
	
	%\bibliographystyle{IEEEtran}

	%%%%%%%%%%%%%%%%%%%%%%%%%%%%%%%%%%%%%%%%%%%%%%%%%%%%%%%%%%%%%%%%%%%%%%%%%%%%%%%%
	%%%%%%%%%%%%%%%%%%%%%%%%%%%%%%%%%%%%%%%%%%%%%%%%%%%%%%%%%%%%%%%%%%%%%%%%%%%%%%%%
	%% Using BibTex
	\bibliographystyle{IEEEtran}
	%\bibliography{IEEEabrv,../MyBibliography}
%	\bibliography{IEEEabrv,MyBibliography}

	%% Or manually put reference below

	%%%%%%%%%%%%%%%%%%%%%%%%%%%%%%%%%%%%%%%%%%%%%%%%%%%%%%%%%%%%%%%%%%%%%%%%%%%%%%%%
	%%%%%%%%%%%%%%%%%%%%%%%%%%%%%%%%%%%%%%%%%%%%%%%%%%%%%%%%%%%%%%%%%%%%%%%%%%%%%%%%
	%% Biography

	%%%%%%%%%%%%%%%%%%%%%%%%%%%%%%%%%%%%%%%%%%%%%%%%%%%%%%%%%%%%%%%%%%%%%%%%%%%%%%%%
	%%%%%%%%%%%%%%%%%%%%%%%%%%%%%%%%%%%%%%%%%%%%%%%%%%%%%%%%%%%%%%%%%%%%%%%%%%%%%%%%
\end{document}